\documentclass{ptephy_v1}
\preprintnumber{XXXX-XXXX} 

\usepackage{natbib} 
\usepackage{url} 
\usepackage{lineno}

\usepackage{ulem}
\begin{document}
\title{Performance of the KAGRA detector during the first joint observation with GEO\,600 (O3GK)}

\author[1]{H.~Abe}
\author[2]{R.~X.~Adhikari}
\author[3,4]{T.~Akutsu}
\author[5,6]{M.~Ando}
\author[7]{A.~Araya}
\author[3]{N.~Aritomi}
\author[8]{H.~Asada}
\author[9,10]{Y.~Aso}
\author[11]{S.~Bae}
\author[12]{Y.~Bae}
% ...

\affil[1]{Graduate School of Science, Tokyo Institute of Technology, 2-12-1 Ookayama, Meguro-ku, Tokyo 152-8551, Japan}
\affil[2]{LIGO Laboratory , California Institute of Technology, 1200 East California Boulevard, Pasadena, CA 91125, USA}
\affil[3]{Gravitational Wave Science Project, National Astronomical Observatory of Japan (NAOJ), 2-21-1 Osawa, Mitaka City, Tokyo 181-8588, Japan}
\affil[4]{Advanced Technology Center, National Astronomical Observatory of Japan (NAOJ), 2-21-1 Osawa, Mitaka City, Tokyo 181-8588, Japan}
\affil[5]{Department of Physics, The University of Tokyo, 7-3-1 Hongo, Bunkyo-ku, Tokyo 113-0033, Japan}
\affil[6]{Research Center for the Early Universe (RESCEU), The University of Tokyo, 7-3-1 Hongo, Bunkyo-ku, Tokyo 113-0033, Japan}
\affil[7]{Earthquake Research Institute, The University of Tokyo, 1-1-1 Yayoi, Bunkyo-ku, Tokyo 113-0032, Japan}
\affil[8]{Department of Mathematics and Physics, Hirosaki University, Hirosaki City, Aomori 036-8561, Japan}
\affil[9]{Kamioka Branch, National Astronomical Observatory of Japan (NAOJ), 238 Higashi-Mozumi, Kamioka-cho, Hida City, Gifu 506-1205, Japan}
\affil[10]{The Graduate University for Advanced Studies (SOKENDAI), 2-21-1 Osawa, Mitaka City, Tokyo 181-8588, Japan}
\affil[11]{Korea Institute of Science and Technology Information (KISTI), 245 Daehak-ro, Yuseong-gu, Daejeon 34141, Republic of Korea}
\affil[12]{National Institute for Mathematical Sciences, 70 Yuseong-daero, 1689 Beon-gil, Yuseong-gu, Daejeon 34047, Republic of Korea}
% ...

\collaborator{KAGRA Collaboration}
\date{\today}
\begin{abstract}%
KAGRA, the kilometer-scale underground gravitational-wave detector, is located at Kamioka, Japan. In April 2020, an astrophysics observation was performed at the KAGRA detector in combination with the GEO\,600 detector; this observation operation is called O3GK. The optical configuration in O3GK is based on a power recycled Fabry-P\'{e}rot Michelson interferometer; all the mirrors were set at room temperature. The duty factor of the operation was approximately 53\%, and the strain sensitivity was $3\times10^{-22}$~$/\sqrt{\rm{Hz}}$ at 250~Hz. In addition, the binary-neutron-star (BNS) inspiral range was approximately 0.6~Mpc. The contributions of various noise sources to the sensitivity of O3GK were investigated to understand how the observation range could be improved; this study is called a ``noise budget.'' According to our noise budget, the measured sensitivity could be approximated by adding up the effect of each noise. The sensitivity was dominated by noise from the sensors used for local controls of the vibration isolation systems, acoustic noise, shot noise, and laser frequency noise. Further, other noise sources that did not limit the sensitivity were investigated. This paper provides a detailed account of the KAGRA detector in O3GK including interferometer configuration, status, and noise budget. In addition, strategies for future sensitivity improvements such as hardware upgrades, are discussed.
\end{abstract}

\subjectindex{F32}

\maketitle
\section{Introduction} \label{ptep02_sec1}

In 2015, the era of gravitational-wave (GW) astrophysics began when the Advanced Laser Interferometer Gravitational-Wave Observatory (Advanced LIGO) detectors observed a binary black hole merger \cite{Abbott:2016blz} in their first observing run (O1). In addition, a binary neutron star merger was detected by the Advanced LIGO and Advanced Virgo detectors \cite{TheLIGOScientific:2017qsa} in their second observing run (O2). With improved astrophysical sensitivity, 35 candidate GWs, including neutron star-black hole binary mergers \cite{Abbott:2021} and extremely unbalanced mass binary mergers \cite{Abbott:2020.043015} were discovered in the second half of the third observing run (O3b)~\cite{LIGOScientific:2021djp}.

The KAGRA detector is the world's fourth large-scale GW detector located in Japan~\cite{Akutsu:2018axf, Akutsu:2020his}.
Two distinctive features of KAGRA include the use of cryogenic mirrors to reduce thermal noise and its construction at the underground site to isolate from seismic motions are two distinctive features of KAGRA. 
The project was funded in 2010, and the underground tunnel excavation was completed in 2014~\cite{Uchiyama:2014uza}. In October 2015, significant infrastructure installation was completed. The KAGRA detector commissioning that aimed at the first astronomical observation commenced after the cryogenic test operation in 2018~\cite{Akutsu:2019rba}. The detector noise was reduced by 4.5 orders of magnitude by this commissioning~\cite{Akutsu:2020his}. After the commissioning, the KAGRA detector was continuously operated for the observing run from 8:00 UTC on April 7 to 0:00 UTC on April 21, 2020. Initially, KAGRA had planned to join the LIGO-Virgo O3 observation. However, while KAGRA was still under commission, O3 was suspended on March 27, 2020, because of the COVID-19 pandemic. Fortunately, the GEO\,600 detector~\cite{Dooley:2015fpa, Lough:2020xft} was online when KAGRA initiated the observing run. Thus, we could conduct a joint observation GEO\,600-KAGRA (O3GK).
During O3GK, the binary-neutron-star (BNS) range, which is the observable distance of gravitational wave from binary-neutron-star coalescence~\cite{Finn:1993}, of the detectors KAGRA and GEO\,600 were 0.6~Mpc and 1.0~Mpc in the median, respectively~\cite{O3GKDApaper}.

This study discusses the performance of the KAGRA detector in O3GK, which considers the limiting factors of the sensitivity. It is essential to argue the possible noise reduction strategies for the subsequent observing runs to improve the detector sensitivity to contribute to the astrophysical searches.

\section{Overview of the KAGRA detector configuration during O3GK}
\label{sec:overview}

The optical configuration of the KAGRA detector, sensing and control sequence for the operation, calibration scheme, and operational status during O3GK are explained in this section as an introduction to the noise coupling topic discussed in Section~\ref{ptep02_sec3}.

\subsection{Interferometer configuration during O3GK}
\label{ss:ifo config}

\begin{figure}[!p]\centering
    \includegraphics[width=15cm]{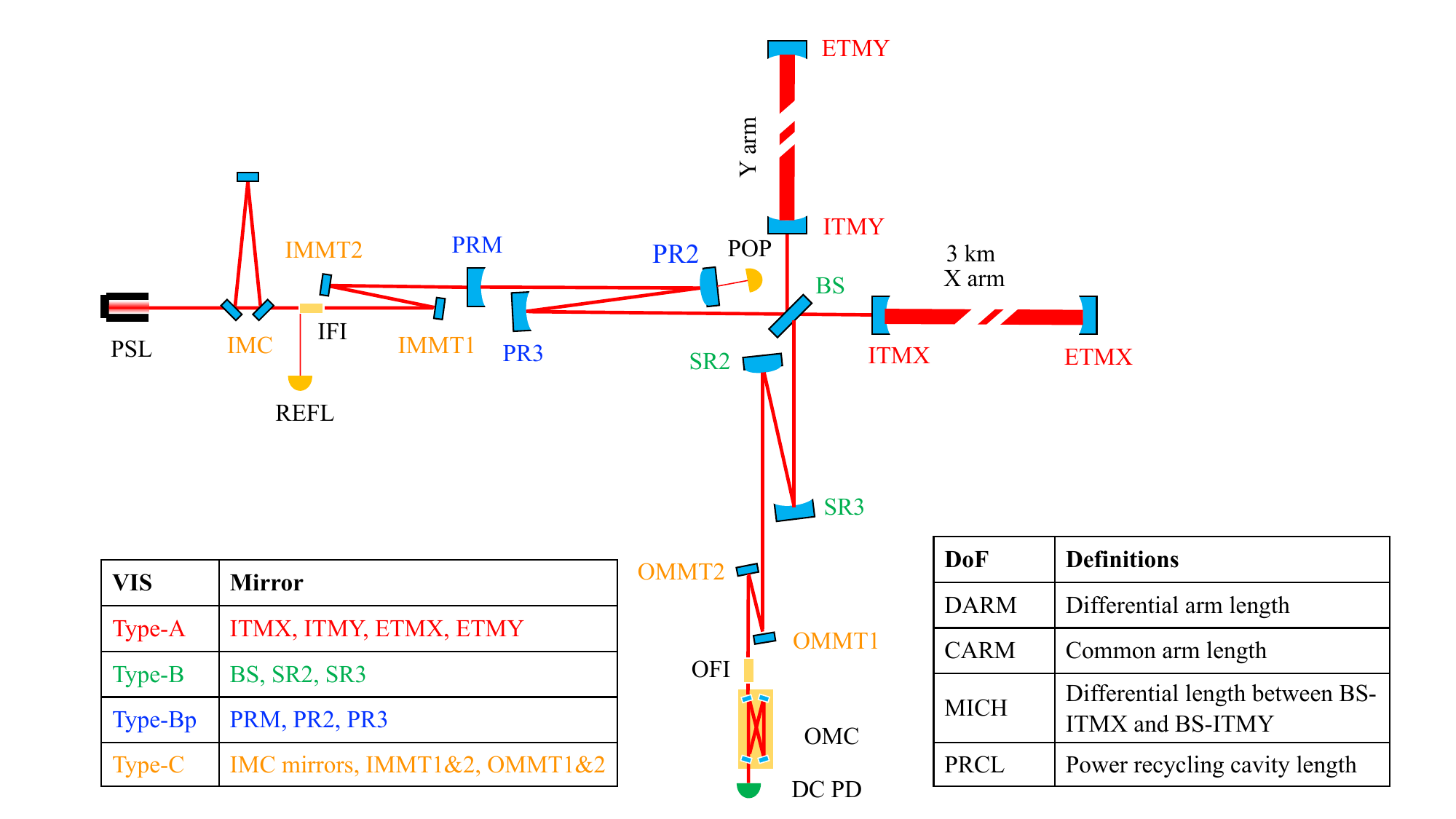}
	\caption{\label{fig:ifo}Schematic of the KAGRA interferometer. A pre-stabilized laser (PSL) acquires phase modulation sidebands at 16.87 MHz and 45.00 MHz in the super-clean PSL room for signal extraction techniques (such as the Pound-Drever-Hall method~\cite{Drever:1983qsr}). Thereafter, using a triangular input-mode cleaner (IMC), the beam is spatially cleaned. An input Faraday isolator (IFI) is used to prevent the laser beam from returning to the PSL from the main interferometer. The input mode-matching telescope (IMMT) mirrors shape the laser spatial mode to match with the main interferometer. The power-recycling mirrors (PRM, PR2, and PR3), beam splitter (BS), and input test masses (ITMX and ITMY) form the power recycling cavity (PRC) to enhance the laser power inside the interferometer. The BS is used to split the beam into two in the X and Y directions. The input test mass X (ITMX), end test mass X (ETMX), input test mass Y (ITMY), and end test mass Y (ETMY) form the 3-km Fabry--P\'{e}rot cavities in the X and Y arms, respectively. The two signal-recycling mirrors (SR2 and SR3) served only as the steering mirrors in O3GK. The SRM (with 30\% power transmittance) was installed at downstream of the SR2; however, it is misaligned to prevent a cavity formation in O3GK, and it is not shown in this figure. The output mode-matching telescopes (OMMTs) match the spatial mode with the output mode-cleaner (OMC) that cleans the spatial mode and rejects unwanted RF sideband fields at the detection port. An output Faraday isolator (OFI) is implemented just upstream of the OMC. The DC photodetector (DC PD) detects the degree of freedom of the differential arm length (DARM), which is a GW channel. The other sensing signals are obtained from three detection ports: reflection (REFL), pick-off port of the power recycling cavity (POP), and anti-symmetric (AS) ports (the DC PD part in this figure). Four length degrees of freedom (right bottom inset) must be controlled to operate the interferometer. The interferometer mirrors are suspended using vibration isolation systems (VISs) to attenuate unwanted movements of the mirrors caused by ground motion. There are four different types of VIS in KAGRA. The left bottom inset summarizes VIS’s and suspended mirrors.}
\end{figure}

The interferometer configuration during O3GK is based on a power-recycled Fabry--P\'{e}rot--Michelson interferometer, as shown in as Fig.~\ref{fig:ifo}. The optical path differences between the two L-shaped orthogonal arms caused by the space--time perturbations are detected at the anti-symmetric (AS) port (the DC PD part in Fig. \ref{fig:ifo}) where the interference fringe is controlled to be dark. In the Fabry--P\'{e}rot cavities, the phase change of the laser field is accumulated for the cavity storage time and enhanced.
Fig. \ref{fig:ifo} shows that the power-recycling mirrors (PRM, PR2, and PR3), beam splitter (BS), and input test masses (ITMX and ITMY) form the power recycling cavity (PRC) that enhances the circulating laser power in the interferometer, and this helps improve the signal-to-noise ratio for the shot noise.
The laser was composed of a non-planar ring oscillator, Mephisto 500NEFC (Coherent), and a fiber laser amplifier, PSFA-10 mw-40 W-1064 (Coherent/Nufern), operating at 1064~nm. The nominal laser power injected to the interferometer (before PRM) was about 5~W during the run.

Optical cavities comprising the complex interferometer need to be controlled to ensure that the cavities are in resonance with the carrier light for operating the power-recycled Fabry--P\'{e}rot--Michelson interferometer as a GW detector. The phase difference between the reflected beams from the two arms also needs to be controlled, and therefore, the AS port is kept dark to ensure the best signal-to-noise ratio for the GW detection.
In total, there are four length degrees of freedom to control: the differential length of the arm cavities and corresponding to the GW degree of freedom (DARM), common length of the arm cavities (CARM), differential motion between the BS and each input test mass (MICH),
and the length of the power-recycling cavity (PRCL). The bottom-right inset in Fig.~\ref{fig:ifo} summarizes all degrees of freedom.

Moreover, the sensing and control schemes described in Ref.~\cite{Aso:2013eba} are applied. Radio frequency (RF) sideband fields at two different frequencies (16.87~MHz and 45.00~MHz) are generated by electro-optic modulators in the pre-stabilized laser (PSL) system to determine the multiple length degrees of freedom of the interferometer \cite{Yamamoto:2019fcc}. In addition, the modulation for the length sensing of the IMC (13.78 MHz) is applied. The light fields are detected and demodulated at the REFL, POP, and AS ports. Signals sensitive to a specific degree of freedom are used as error signals for the feedback control loops.
The GW signals (or DARM signals) are detected as slight DC power changes at the OMC transmission port known as the DC readout scheme~\cite{Fricke:2011dv}. 
The OMC is a bow-tie shape optical cavity to clean the spatial mode
and it rejects unwanted RF sideband fields at the AS port.
The parameter sets of the optical properties are summarized in Table \ref{table:param}.

The interferometer mirrors are suspended using vibration isolation systems (VISs) composed of multiple mechanical stages to attenuate unwanted movements of the mirrors caused by ground motion.
As mentioned in the inset of Fig. \ref{fig:ifo}, there are four different types of VIS in KAGRA.
The test masses (ETMX, ETMY, ITMX, and ITMY) are suspended by Type-A suspensions that are 13-m tall 9-stage pendulums. The first five stages are called Type-A towers, which are at room temperature~\cite{Okutomi:2019, Fujii:2020}; and the last four stages are called cryopayload, which can be cooled down~\cite{Akutsu:2020wgy}. 
The BS, SR2, and SR3 are suspended by Type-B suspensions, which have a two-stage payload under a three-stage seismic isolation system that includes inverted pendulums \cite{Akutsu:2021auw}.
The PRM, PR2, and PR3 are suspended by Type-Bp suspensions and are similar to Type-B suspensions except without the inverted pendulums \cite{Akiyama:2019ycw}. 
Small auxiliary mirrors (IMC mirrors, IMMT1,~2, and OMMT1,~2) are suspended by Type-C suspensions, which are simple double-pendulums originally designed for TAMA300 detector~\cite{Takahashi:2002cs}.
Moreover, the aluminum breadboard of the OMC cavity was suspended by three blade springs made of maraging steel in O3GK.

One key feature of KAGRA to reduce thermal noise is the cryogenic technique \cite{Akutsu:2020his, Akutsu:2019rba} to cool the cryopayloads. Before O3GK, all cryogenic systems were already installed. The cryopayloads were connected to the pulse tube cryocoolers by soft heat links made of pure aluminum to conduct the heat from the cryopayloads to the cryocolers. Further, additional VISs specially designed for the heat links were installed to avoid transferring ground vibrations. However, the cryopayloads were at room temperature during O3GK because the sensitivity was not yet limited by mirror thermal noise. These cryocoolers will operate in future observations, and the sapphire mirror (TM) temperature will be approximately 20 K.

The suspension controls (Sections \ref{ss:Type A sensor} and \ref{ss:Type B sensor}) and interferometer controls (Section \ref{sec:lock}) are performed using the real-time digital control system originally developed by LIGO. Analog signals from the instruments are converted to digital signals using the analog-to-digital converter (ADC) cards, and they are filtered using digital servo filters. The digital control signals are converted to analog signals using the digital-to-analog converter (DAC) cards and fed back to the relevant actuators.
Control loops that require broad control bandwidths are performed in analog, whereas the digital system continue to obtain the monitoring signals.
Furthermore, the real-time digital control system is integrated into the data acquisition system, and together, they are called the control and data acquisition system (CDS)~\cite{Bork:2020zet}.

\begin{table}[!t]\centering
\begin{tabular}{|l|c|r|}
\hline
Length & IMC & 53.30 m \\ \cline{2-3}
& Power-recycling cavity &  66(1) m\\ \cline{2-3}
& X arm & 3000 m \\ \cline{2-3}
& Y arm & 3000 m \\ \cline{2-3}
& Schnupp asymmetry & 3.36(1) m\\ \hline \hline
RF sideband frequency & IMC modulation & 13.78 MHz \\ \cline{2-3}
& 1st modulation & 16.87 MHz \\ \cline{2-3}
& 2nd modulation & 45.00 MHz \\ \hline \hline
Modulation index & IMC modulation & not measured \\ \cline{2-3}
& 1st modulation & 0.23(2) rad \\ \cline{2-3}
& 2nd modulation & 0.23(2) rad \\ \hline \hline
Power recycling gain & \multicolumn{2}{r|}{11(1)} \\ \hline \hline
Arm cavity finesse & X arm & 1410(30)  \\ \cline{2-3}
& Y arm & 1320(30)  \\ \hline 
\end{tabular}
\caption{\label{table:param}Measured parameters of the KAGRA interferometer during O3GK. The test masses were maintained at room temperature. }
\end{table}

\subsection{Lock acquisition}
\label{sec:lock}
Lock acquisition is a control sequence that simultaneously achieves the resonances of all cavities in the interferometer and the dark fringe at the AS port. The mirrors move larger than resonance widths of cavities, and therefore, each cavity must be controlled within its linear regime (where it is said ``the cavity is {\it locked}'') because the responses of the light fields strongly depend on the resonant conditions of the cavities in the interferometer. 
The laser interferometer can be operated as a GW detector only when all length degrees of freedom are locked.

The lock acquisition scheme of the KAGRA detector is similar to the Advanced LIGO scheme~\cite{Staley:2014csa} but with two differences. First, there is no signal-recycling cavity; therefore, there is one less degree of freedom to control. Second, the optical configuration of the arm length stabilization (ALS) system is different. The ALS system utilizes an auxiliary green visible laser that is phase locked to the PSL light with a doubled frequency of the PSL light; this green laser independently controls the arm cavities from the resonant conditions of the main IR laser. 
As described in Ref.~\cite{Akutsu:2019gxm}, in KAGRA, the green laser beams are injected through the backside of the PR2 and SR2 mirrors. In LIGO, the green laser beams go into the arm cavities through the ETMs.

The main optics of the interferometer are initially aligned before the lock acquisition procedure, as follows. Each arm cavity is independently aligned such that the high-visibility flashes of green light are observed by the PD at each transmission port and the CCD camera looking at each end mirror surface~\cite{Akutsu:2020tpd}, whereas all the other mirrors are misaligned. After recording the optimal angles of ETMX, ITMX, ETMY, and ITMY mirrors, the short Michelson interferometer (composed of the BS, ITMX, and ITMY mirrors) is configured, and the BS is aligned to attain the high visibility of the Michelson interferometer. Then, the PRM is aligned in the PRC configuration.

The PRM is misaligned after recording its optimized angle by confirming the PRC flashes; the interferometer is configured as a Fabry--P\'{e}rot--Michelson interferometer.
In this configuration, the arm cavities are maintained non-resonant for the main IR beam by the ALS system. The common arm signals of the ALS (ALS CARM) and the differential arm signals (ALS DARM) are fed back to the frequency of the main laser to which the green laser is phase locked and the differential motion of the end test masses (ETMX and ETMY), respectively.
The ALS system adds an offset to the ALS CARM such that the arm cavities are at off resonance. 
This scheme is useful because the sign of the carrier fields in the PRC and the short Michelson interferometer section (the central part) depends on the resonant condition of the arm cavities. The sign of the carrier field does not switch because of maintaining the arm cavities non-resonant for the carrier field; therefore, the signs of the PRCL and MICH length signals do not switch during the lock acquisition of the central part of the interferometer.

The PRM is subsequently realigned to configure the power-recycled Fabry--P\'{e}rot--Michelson interferometer with non-resonant Fabry--P\'{e}rot cavities. The PRCL and MICH are locked using the third harmonic demodulation signals ({\it ``3f signals''}) that are less sensitive to the sign switch of the carrier field \cite{Arai:2000ue}.
The incident power to the arm cavities are stabilized when the MICH and PRCL degrees of freedom are locked by the $3f$ signals. With the MICH and PRCL being locked by the $3f$ signals, the arm cavities are brought to their resonances by reducing the offsets on ALS CARM and DARM signals; then, the control signals are transitioned to the Pound-Drever-Hall (PDH) \cite{Drever:1983qsr} signals obtained at REFL and AS ports for CARM and DARM, respectively.

Finally, the DARM length signal is switched from the demodulated signal to the DC readout signal with the DARM offset of about 17 pm. 
The error signal of the OMC cavity length is obtained by dithering the OMC length and demodulating the detected signal, and by feeding it back to the PZT attached to one of the four mirrors of the OMC.

The entire locking procedure is automated using the state machine Guardian~\cite{GraefRollins2016xhy} developed by the Advanced LIGO project. This locking sequence takes about 20-30~min depending on exactitude of the initial alignment and on environmental conditions such as the degree of ground motions. Further details regarding control scheme will be presented in our future paper on the detector commissioning. 

\subsection{Alignment control}
\label{sec:asc}
The alignment degrees of freedom for the interferometer must be controlled~\cite{Barsotti:2010zz} in addition to the length degrees of freedom (Sec.~\ref{sec:lock}).
Mirror misalignment is gradually induced by external disturbances such as ground motions, and this affects the long-term stability and noise coupling for the DARM degree of freedom.
The following three alignment sensing and control schemes are used in GW detectors:
RF wavefront sensors~\cite{Fritschel:1998ctz},  alignment dithering~\cite{Kawabe94}, and beam spot centering. In O3GK, an alignment dither system (ADS) was implemented for local alignment control. The ADS method uses mechanical modulation in the mirror's angular degrees of freedom and the lock-in detection.

During O3GK, PR3, BS, and IMMT2 were intentionally excited in angular degrees of freedom ({\it i.e.}, in pitch and yaw). The RF signal at the POP port was demodulated at 90~MHz to obtain the build-up of the 45~MHz sideband field in the PRC; this was again demodulated at the angular excitation frequencies (lock-in detection).
Thus, ADS loops were used to maximize the buildup of the 45~MHz sideband; consequently, the carrier and 16~MHz sideband fields were also maximized in the PRC.

\subsection{Calibration and sensitivity}
Detector sensitivity in displacement is reconstructed by calibrating detector outputs in the DARM control loop using the interferometer response and coil-magnet actuator functions. The details of the O3GK calibration are described in Refs.~\cite{O3GKDApaper, Akutsu:2020tpd}.
The calibrated data used in this study are derived in a real-time digital control system with infinite impulse response filters that simulate the pre-measured optical response and actuator functions with an error of approximately 15\% in magnitude. However, this calibration does not consider any time-dependent factors of the interferometer response induced by, for example, alignment drifts, because this calibration pipeline aims to realize only detector commissioning and monitoring.

\begin{figure}[!t]\centering
    \includegraphics[scale=0.4]{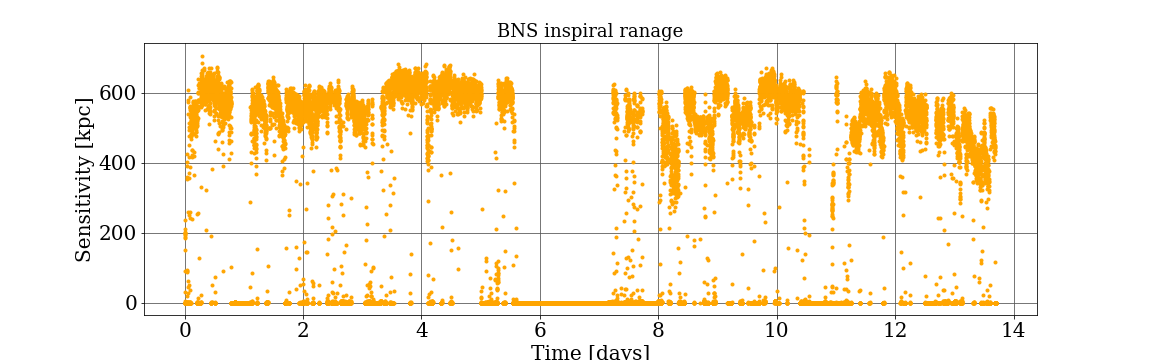}
    \caption{\label{fig:BNS} BNS inspiral range of the KAGRA detector during O3GK, from 8:00 UTC on April 7, 2020. The median BNS inspiral range during the observation was 0.6~Mpc. The detector was non-operational during the latter part of days 5 to day 7 because of a storm.}
\end{figure}

Using the calibrated data, astrophysical BNS inspiral ranges were calculated, as depicted in Fig.~\ref{fig:BNS}. The BNS inspiral range represents the detection range of the volume- and orientation-averaged inspiral signals from a 1.4 solar mass BNS inspiral at a signal-to-noise ratio of 8~\cite{Finn:1993}.
The presented BNS inspiral range was calculated during the observation, using the calibrated data derived from the above-mentioned method without considering any redshift effect because the expected BNS inspiral range was less than 1 Mpc.

\subsection{Duty factor}
\label{sec:duty}
The interferometer occasionally experiences {\it lockloss} wherein the control loops can no longer maintain the cavity resonances. This is attributed to various factors such as significant seismic activities and global misalignment caused by interferometer mirrors drifting away over time. The global misalignment is believed to cause also sensitivity degradation even if it does not break the lock; however, it is still under investigation.

The KAGRA detector was operational for approximately 53\% of the entire time of the observation run time. This duty factor of GEO\,600 was 78\%, which results in a 45\% coincident time of the two detectors.

As depicted in Fig.~\ref{fig:BNS}, the interferometer was not operational around day 6 and day 7 because of the high seismic activity at 0.1--0.3~Hz bandwidth, known as the micro-seismic motion. For KAGRA, the micro-seismic activity is excited by the waves in the sea of Japan and the Pacific Ocean. During day 2, the micro-seismic motion was as high as 1--2~$\mu$m/s~\cite{1857988}. The other occasional downtime was caused by earthquakes, abnormal behavior of the control loops, and gradual drifts of the interferometer alignment. Once the interferometer lost control, the entire lock acquisition sequence was repeated automatically by the Guardian. The initial alignment was manually performed before the automatic lock acquisition depending on the interferometer condition (for instance, when the visibility of the short Michelson interferometer was degraded).

\begin{figure}[!t]\centering
    \includegraphics[scale=0.4]{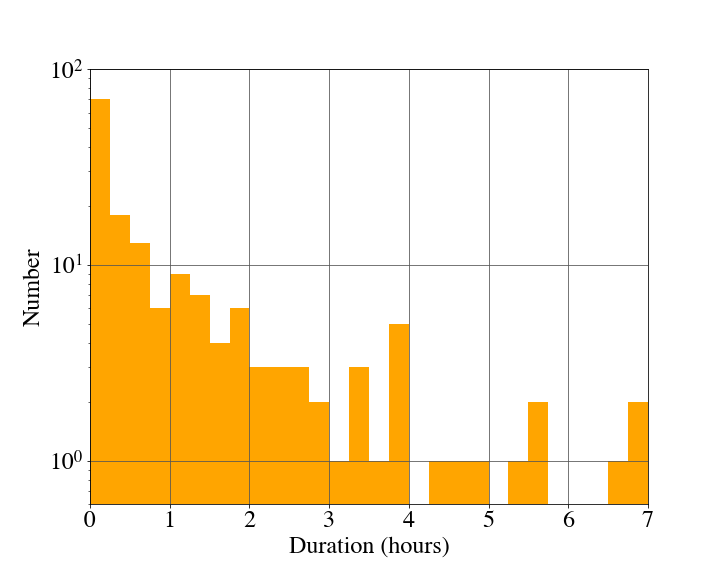}
    \caption{\label{fig:hist} Histogram of the operational duration during O3GK.}
\end{figure}

As depicted in Fig.~\ref{fig:hist}, a majority of the observation segments had an operational time duration of less than 15 min. They correspond to about 10\% of the total operational time. We expect to improve the operational duration once the global angular control system with wavefront sensors is implemented.

\section{Noise Budget} \label{ptep02_sec3}
\begin{figure}[!p]\centering
    \includegraphics[width=15cm]{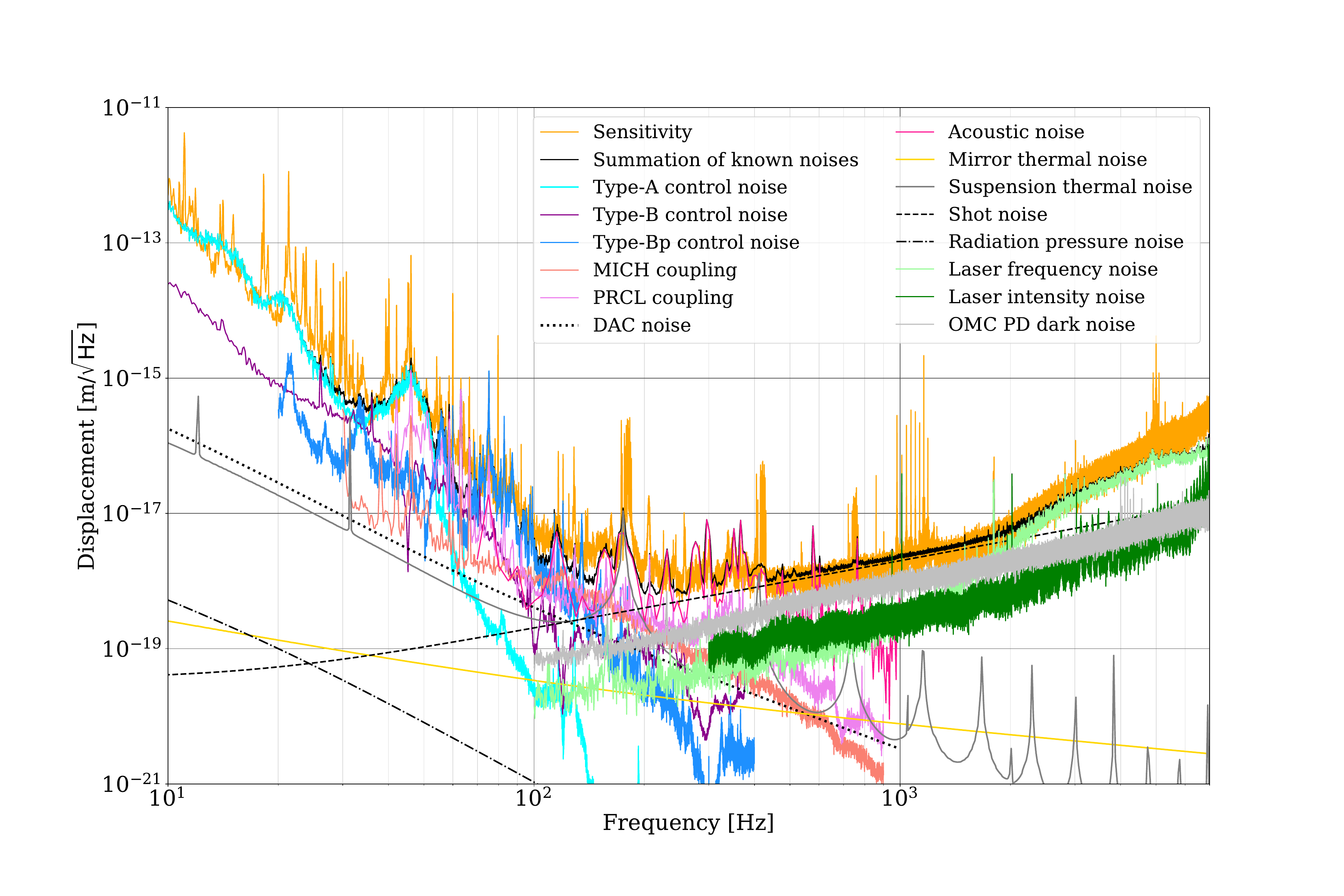}
    \includegraphics[width=15cm]{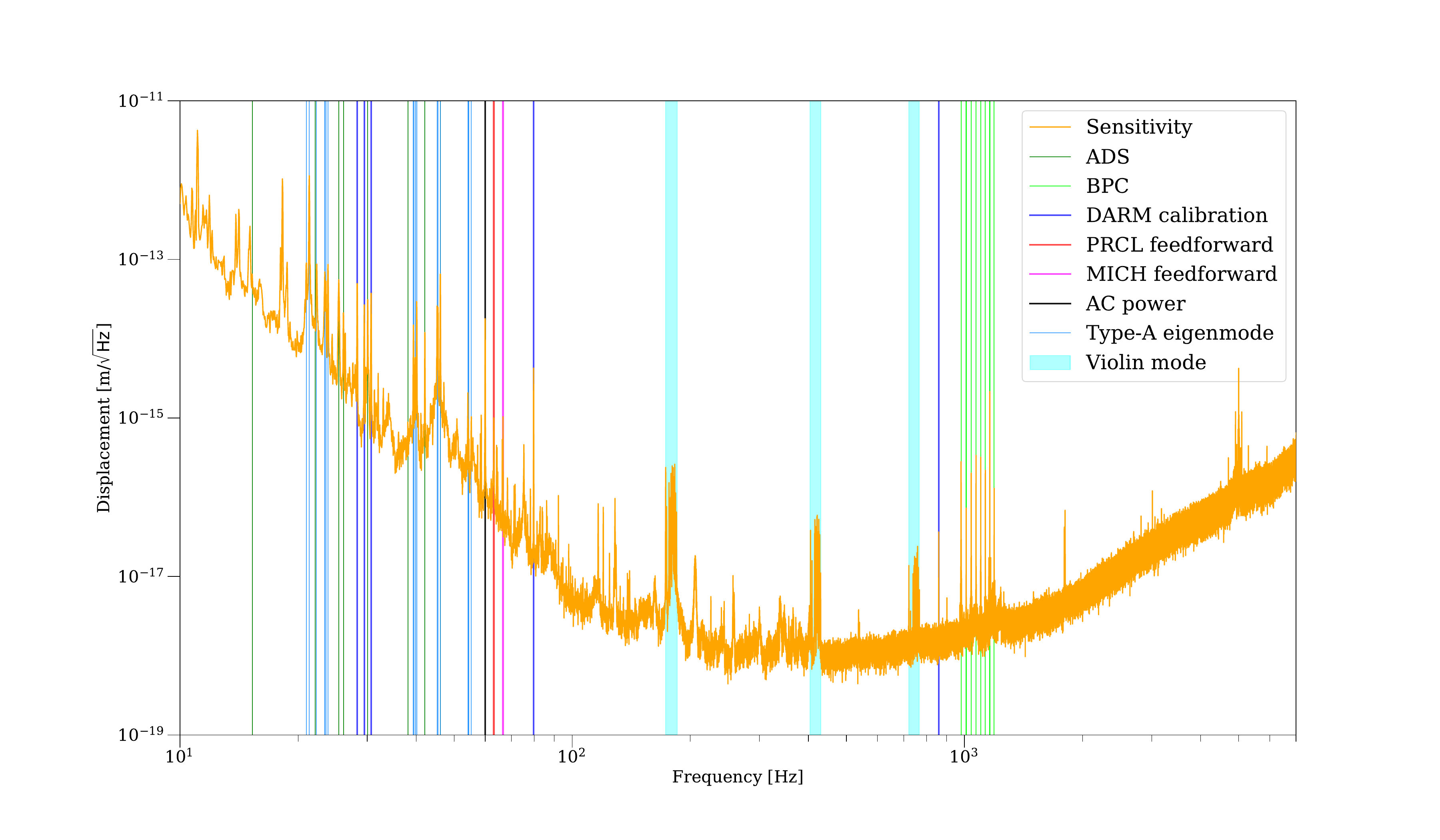}
    \caption{Top: Various noise sources in the KAGRA detector during the O3GK run (April 12th, 2020). The measured sensitivity (orange curve) and summation of various noise sources (solid black curves) are displayed. Each noise source (depicted in multiple colors) is estimated using online monitoring channels and pre-measured transfer functions.
    Bottom: Measured sensitivity (orange curve) and known line frequencies (vertical lines). The values of each frequency are listed in the Appendix~\ref{ptep02_linelist}.\label{fig:NB}}
\end{figure}

For noise reduction, it is necessary to evaluate the noise sources and their couplings; this study is called ``{\it noise budget}.'' KAGRA has already provided noise budgets for the first operation using a simple Michelson interferometer configuration in March and April 2016~\cite{Akutsu:2017kpk}, and, for the first cryogenic operation in April and May 2018~\cite{Akutsu:2019rba}. A similar study was repeated for the power-recycled Michelson interferometer configuration at room temperature of O3GK. Each noise contribution in the noise budget (top panel of Fig.~\ref{fig:NB}) and the reduction strategies are explained in the following sections from lower to higher frequencies. In each frequency region, the largest contribution is displayed primarily, and the identified peaks (bottom panel of Fig.~\ref{fig:NB}) are also explained. The measured sensitivity can be explained by adding the effects of each noise.

The estimated noise curves in the sensitivity are evaluated as follows: the amplitude spectral density of the witness channel for a certain noise contribution and the transfer function from this witness channel to the GW channel (DARM) are measured. For the latter, the sinusoidal signal is injected intentionally, and the ratio of the Fourier components at the DARM to the witness channel is evaluated. In some cases, the simulated transfer functions or the theoretical formula of the noise contribution are adopted.

\subsection{Noise contribution below 50~Hz}

Below 50~Hz, the noise of the photocoupler position sensors for Type-A suspension's local control limited the sensitivity. The other noise contributions (seismic motion through Type-A tower and vibrations through the heat links) and the resonant peaks of the Type-A suspension are also discussed. Although Newtonian noise can be one of the limiting noise sources for the terrestrial detectors in this frequency region, it is not discussed here because it is sufficiently smaller than the KAGRA design sensitivity owing to the underground site~\cite{Badaracco:2021}.

\subsubsection{Control noise of the cryopayload part of the Type-A suspensions}\label{ss:Type A sensor}

All mirrors shown in Fig.~\ref{fig:ifo} are isolated from the seismic motion in the GW frequency band because of the VISs. In contrast, around the resonant frequencies of the suspensions of VISs, which are lower than the GW frequency, the mechanical resonances enhance the vibration. Local damping controls were implemented to suppress the vibration of the suspensions. The local sensors monitor the internal vibration, relative motions between suspension parts or motion relative to the ground. The sensor signals are used to reduce the resonant motion via appropriate feedback loops.

\begin{figure}[!t]\centering
    \includegraphics[width=15cm]{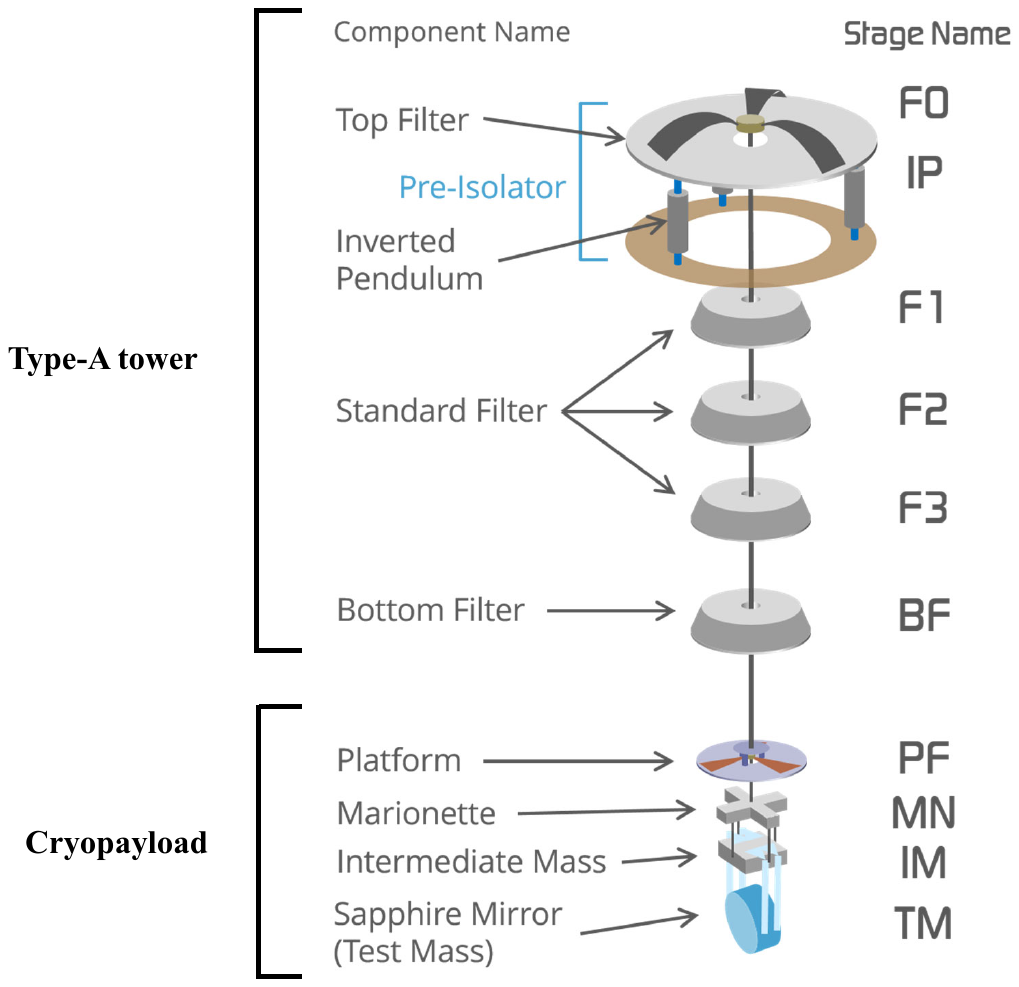}
	\caption{\label{fig:TypeA}Schematic of Type-A suspension~\cite{Okutomi:2019,Fujii:2020}. The suspension is a multiple-stage pendulum with a total height of 13~m. The Type-A tower is at room temperature while the cryopayload is cooled down to cryogenic temperature. The foot of the inverted pendulum for horizontal vibration isolation is fixed to the ground, and it suspends a four-stage pendulum at room temperature. For vertical vibration isolation, each inverted pendulum and suspension stage has geometric anti-spring (GAS) filters~\cite{Bertolini:1999}.  The bottom of this pendulum, the bottom filter (BF), suspends four stages of the pendulum at the cryogenic temperature. The platform at the top of the cryopayload has leaf springs for the vertical vibration isolation, and it suspends the marionette (MN), intermediate mass (IM), and sapphire mirror (TM). The IM has four sapphire blade springs for vertical vibration isolation. The photocoupler position sensors (``{\it photo-sensors}'') to monitor the internal vibration resonances of the cryopayload were implemented as the local damping sensors \cite{Akutsu:2020wgy}. Coil magnet actuators are installed to apply feedback force on the MN, IM, and TM for local damping and lock acquisition. The damping filters were implemented to reduce the free-swinging motion in the inverted pendulum, GAS filters, BF, MN, and IM. Further, the sensor correction system was implemented to reduce the contamination of the seismic noise at the inverted pendulum stage \cite{Fujii:2020}.}
\end{figure}

A simplified drawing of the Type-A suspension is depicted in Fig.~\ref{fig:TypeA}. For the Type-A suspensions, the feedback force is applied to the Marionette (MN), intermediate-mass (IM), and sapphire mirror (TM) with the coil magnet actuators.
In this scheme, the sensor noise applied to the suspensions through the feedback system contaminates the GW signals. The local damping for the longitudinal direction (which is the optical axis) on the mirror, IM, and MN stages was intended to be used only during the lock-acquisition phase and to be turned off in the observation mode to avoid sensor noise coupling. However, in O3GK, local damping was applied to the MN stage to reduce the residual longitudinal motion of the mirrors.

In O3GK, the photocoupler position sensors (``{\it photo-sensors}'') of cryopayload were adopted as the local damping sensors \cite{Akutsu:2020wgy}. These sensors monitored the internal vibration resonances of the cryopayload. The feedback signals were sent to the actuators to control the position and angle of the MN. 

A transfer function from the feedback signal of the MN stage to the DARM was measured before O3GK. The photo-sensor noise contribution to the DARM was estimated using the online witness channel of the MN photo-sensor in the O3GK observation; it was multiplied by the transfer function. We observed that the amplitude of each TM motion caused by the photo-sensor noise was different in the four test masses in O3GK. Among the four, the photo-sensor for ITMX was the noisiest, and it limited the O3GK sensitivity to below 50 Hz.

Although local sensors would not be used for longitudinal motion in the final phase of the KAGRA project where the sensitivity is considerably better, they will still be necessary in the near future.
We plan to substitute the photo-sensors with optical levers for MN damping to sense the motion relative to the ground; the investigation is in progress. Another method for improvement is to increase the sensitivity of the photo-sensors by modifying the driving and readout circuits. With improved photo-sensors, it is essential to modify the damping loop to reduce the feedback gain at higher frequencies (i.e., in the GW band) to lower the sensor noise coupling.

\subsubsection{Seismic noise through Type-A tower}
As a unique key feature, the KAGRA detector is located in a 200-m deep underground site where the seismic motion at the GW frequency is reduced by approximately $1/100$ compared to the surface \cite{Akutsu:2020his}.

In addition, the VISs were installed to isolate the main KAGRA mirrors from the seismic motion.  
They isolate the vibration of the mirrors in the horizontal (along the laser beam) and vertical directions of the mirrors. The laser beam is not exactly perpendicular to the vertical direction because the arm baselines have a slope inclination of $1/300$ for an underground water drain. Moreover, imperfections in the vibration isolation suspensions convert the vertical motions into horizontal motions. The actual coupling factor was measured to be 1\% from the vertical to horizontal motions at the mirror stage of Type-A suspensions~\cite{Shoda:2020}. Multiple stage pendulums and an inverted pendulum are adopted for horizontal vibration isolation. The geometric anti-spring (GAS) filters are deemed effective for vertical vibration isolation~\cite{Bertolini:1999}. For Type-A suspensions, each pendulum stage at the room temperature has a GAS filter, and in total, there are five GAS filters in a Type-A tower.

\begin{figure}[!h]\centering
    \includegraphics[width=7.5cm]{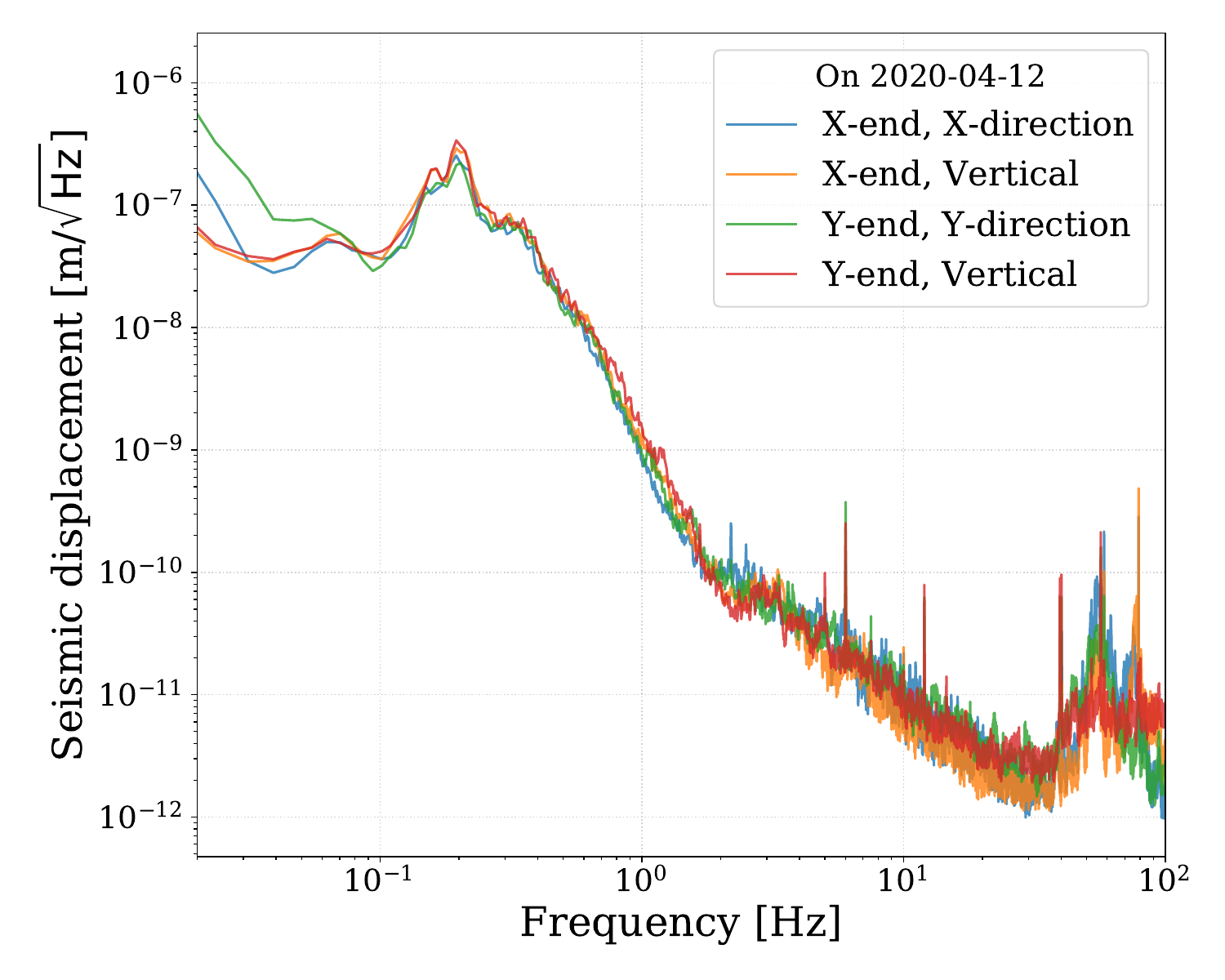}
    \includegraphics[width=7.5cm]{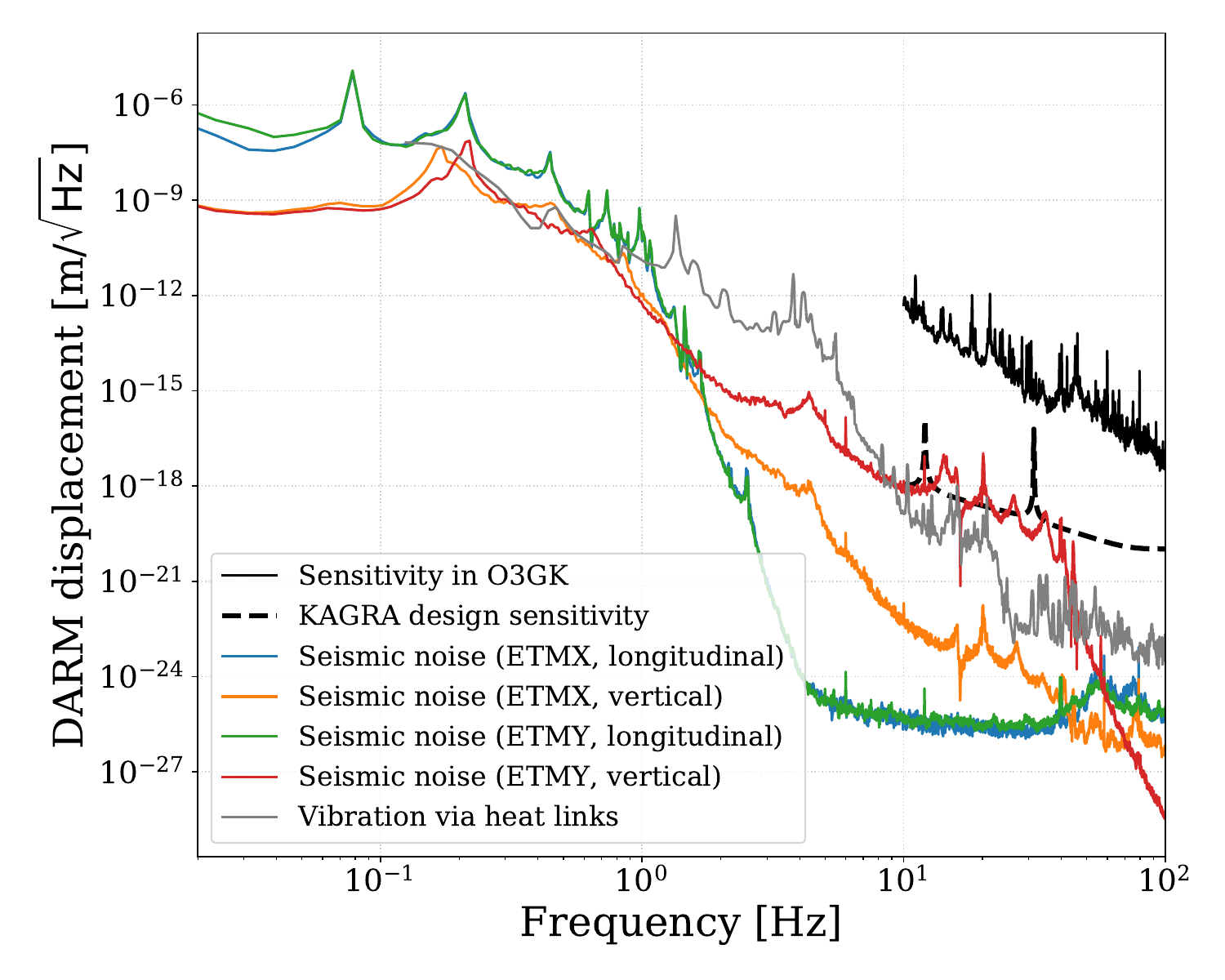}
    \caption{(left) Amplitude spectral density of seismic motion measured by seismometers in the KAGRA site on April 12th, 2020. 
    (right) Estimated noise contribution of seismic motion through the Type-A tower and the vibration through the heat links. The coupling from the vertical seismic motions to the longitudinal direction was assumed to be 1\%. These noise contributions are lower than the sensitivity in O3GK, as indicated in the right inset. Although seismic motions through the ETMY Type-A tower are higher than the design sensitivity, they reduce after restoration. Vibrations through the heat links are sufficiently low because of the heat link vibration isolation system (HLVIS).
    \label{fig:seis_tower}}
\end{figure}

The previously explained seismic noise through the Type-A tower was estimated as follows: the ground motions measured using the three-axis seismometers placed near the ETMX and ETMY in O3GK~\cite{Akutsu:2020tpd} are multiplied by the modeled transfer functions from the ground to the mirror (ETMX or ETMY). 
All Type-A towers had some parts that did not work in O3GK, among which ETMY was extremely ineffective because three GAS filters were stuck. In contrast, the ETMX Type-A tower was effective.
We calculate the seismic noise under these conditions; the results are displayed in Fig.~\ref{fig:seis_tower}.
The seismic noise in O3GK was dominated by the vertical vibration of the ETMY Type-A tower. The seismic contribution of ETMX Type-A tower already met the KAGRA design goal in O3GK, and as such, the seismic contribution of Type-A suspensions is expected to be lower than the KAGRA design sensitivity once the ongoing repairs of Type-A suspensions are complete.

\subsubsection{Vibration through heat links}
Cryogenic mirrors are the key feature of the KAGRA design for reducing thermal noise. Although these mirrors are not cooled during O3GK, heat links (made of aluminum thin wires) are installed to connect the cryocooler and cryopayload for future cooling. Heat link vibration isolation system (HLVIS) was introduced to avoid these links from propagating the vibrations because of seismic motion, cryocoolers, or compressors~\cite{Yamada:2021}.

The noise projection depicted in Fig.~\ref{fig:seis_tower} is calculated as the product of a measured vibration of the radiation shield\footnote{This vibration was measured about 2.5~years before O3GK when the cryostat was evacuated in room temperature, and all cryocoolers were in operation.} and a simulated transfer function from the radiation shield to the TM through an HLVIS. This vibration projection is lower than the KAGRA design sensitivity. 

\subsubsection{Eigenmodes of the Type-A suspensions}
The resonances of the Type-A suspension eigenmodes appeared in the observed sensitivity as various peaks.
Only identified peaks are listed in this paper; however, some unknown lines are also observed. Not all expected peaks are observed, and this is most likely because the background noise is larger than the peaks at some frequencies. In this section, eigenmodes below approximately 50~Hz are discussed. The violin modes of Type-A suspensions, above 150~Hz, are discussed in Sec.~\ref{sec:violin}.

Local damping systems were implemented to suppress the resonant motions; however, they were not perfectly eliminated.
The TM, IM, and MN stages were intentionally (and independently) excited through the corresponding actuators and monitored using the local sensors to identify the resonant frequencies of Type-A suspensions. The results of these measurements were compared with a mathematical model of a Type-A suspension~\cite{Okutomi:2016}.

As depicted in the bottom plot of Fig.~\ref{fig:NB}, five peaks at approximately 21~Hz, 23~Hz, 40~Hz, 45~Hz, and 55~Hz constitute the eigenmodes of the Type-A suspensions. The peak frequencies are listed in Appendix~\ref{ptep02_linelist}.
These resonant peak excitations can be caused by the couplings of the degrees of freedom in a servo control system (for example, longitudinal motion and pitch, yaw rotation); the decoupling (reduction of coupling) is necessary.

\subsection{Noise contribution between 50~Hz and 100~Hz}

The noise of the local damping control in Type-Bp suspensions between 50~Hz and 100~Hz was dominant in the sensitivity. The other noise sources in this frequency range included the control noise from Type-B suspensions, cross-coupled control noise from auxiliary length degrees of freedom, and the electronic noise from DACs and coil drivers, and the mains AC power supply.

\subsubsection{Control noise of Type-B/Bp suspensions}\label{ss:Type B sensor} 

\begin{figure}[!h]\centering
    \includegraphics[width=15cm]{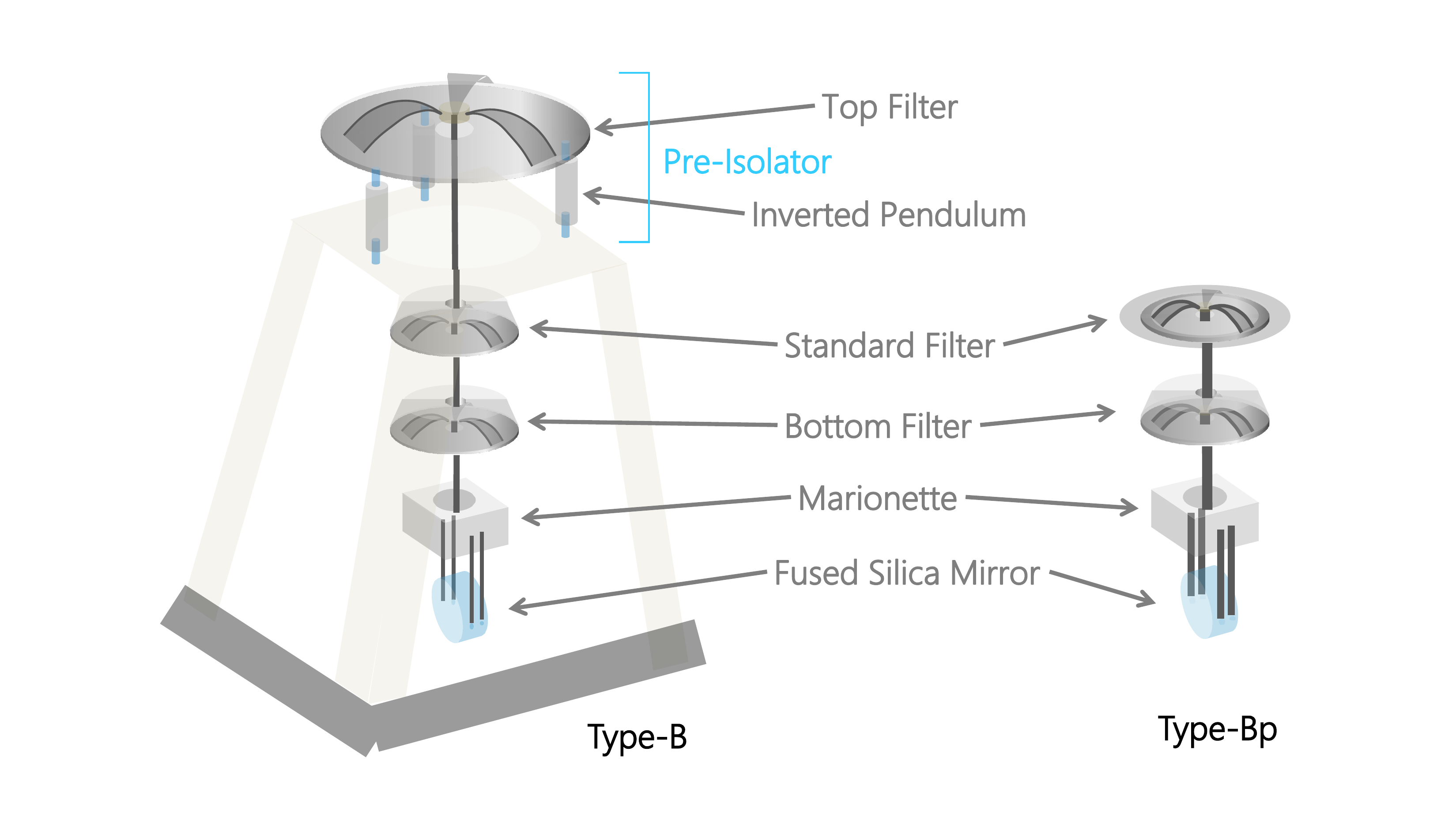}
    \caption{\label{fig:TypeB/Bp}Schematic of Type-B (left) and Type-Bp (right) suspensions. Both suspensions comprise a room-temperature payload and GAS filters for vertical isolation. Further, the Type-B suspension has an inverted pendulum for the horizontal isolation. The payload contains the MN, recoil mass (not shown in the figure), and fused silica mirror. 
    Coil magnet actuators are installed to apply a feedback force on the mirror stage. Optical levers are equipped on the mirror stage to monitor the pitch, yaw, and longitudinal motion of the mirror. Details of Type-B and -Bp suspensions are presented in Ref.~\cite{Akutsu:2021auw} and Ref.~\cite{Akiyama:2019ycw}, respectively.}
\end{figure}

Simplified schematics of Type-B and \mbox{-Bp} suspensions are shown in Fig.~\ref{fig:TypeB/Bp}. Like the Type-A suspensions, the largest noise contribution from the Type-B and \mbox{-Bp} suspensions to the sensitivity included the control noise from the local damping loops on the mirror stage. For Type-B and -Bp, the local sensors used for the control included optical levers for measuring the displacement of the position and angle to ground, in contrast to the photo-sensors that monitor the internal motions of the Type-A suspensions.

The transfer functions from the feedback signals of the mirror stages to the DARM were measured before O3GK to calculate the noise contribution. The local control noise contribution of the Type-B/Bp suspensions to the DARM was estimated using the feedback signals measured during O3GK, and it was multiplied by the transfer functions. 
Only BS was considered for the noise budget for Type-B suspensions because SRM was misaligned and SR2 and SR3 were used only as steering mirrors. The noise contribution from the Type-Bp suspensions includes the three power-recycling mirrors. For Type-B, the contributions along the longitudinal, pitch, and yaw directions were added in the quadrature. For Type-Bp, only the contributions from pitch and yaw were added because the contribution along the longitudinal direction was negligible. 

The top panel of Fig.~\ref{fig:NB} represents the contributions from the Type-B suspension (purple) and Type-Bp suspensions (blue). The noise contribution from the Type-Bp suspensions was larger than that from the Type-B suspension above 50~Hz; the control noise from the Type-Bp suspensions limits the sensitivity between 60~Hz to 100~Hz. 

Control filters must be redesigned to minimize the noise for future improvements. One strategy is to use a high-gain band-pass filters only near mechanical resonances.

\subsubsection{Coupling from the auxiliary length degrees of freedom}
\label{sec:MICH/PRCL noise}

The power-recycled Fabry--P\'{e}rot--Michelson interferometer has four length degrees of freedom as shown in Fig.~\ref{fig:ifo}. Although they are ideally independent of each other, a certain amount of coupling exists. Further, there is an inevitable MICH-to-DARM coupling because the DARM error signal is fundamentally sensitive to the MICH degree of freedom because both are the X and Y arm's differential degrees of freedom.
In addition, the PRCL degree of freedom couples to the DARM motion through the interferometer's practical imperfections. The sensing noise of the MICH and PRCL sensors are re-injected to the DARM control loop through the crosstalk.

We assume that the DARM error signal can be represented by the summation
of the {\it real} DARM signal and the sum of the contributions of auxiliary degrees of freedom
through (unmodeled) linear transfer functions from the control loops in an auxiliary degree of freedom to the DARM loop to evaluate the crosstalk between auxiliary degrees of freedom and DARM. The re-injected sensing noise is subtracted using the transfer function from the control point of the auxiliary degree of freedom loop to the DARM error point~\cite{Allen:1999wh, Meadors:2013lja} by the feed-forward method. 
Here, MICH and PRCL control signals were subtracted from the DARM through ITMX (as an actuator), using the pre-measured and fitted corresponding transfer functions.

The real-time feed-forward subtraction successfully improved the sensitivity between 80~Hz and 200~Hz by approximately an order of magnitude. The residual contribution is depicted in salmon and light magenta colors in the top plot of Fig.~\ref{fig:NB}.
With the feed-forward subtraction, the PRCL-to-DARM contribution was no longer directly limited the sensitivity; however, it was still one of the major noise sources between 50 Hz and 80 Hz, and the MICH-to-DARM coupling was slightly lower than the other noise sources in the frequency region. There is a possibility to double-count the Type-B and -Bp's control noise in the MICH and PRCL lines in the top panel of Fig.~\ref{fig:NB}, because some of the MICH and PRCL noise may be from the suspension control noise. In future commissioning, a similar feed-forward online subtraction will be adopted, wherein an additional degree of freedom (the SRC length) will be included.

\subsubsection{DAC and coil driver noise}
Coil magnet actuators are used to control the positions and angles of the suspended mirrors \cite{Michimura:2017}. Control signals from the real-time control system are transmitted to actuators through the DACs and driver circuits. Any noises caused by the DACs and drivers are converted to the force applied by the coil magnet actuator to the suspended mirrors. Only the DAC noise is considered because the noise from the driver circuits is known to be a few magnitude lower than that of the DAC.

Amplitude spectral density at the DAC output with a null DAC input was measured and multiplied by a pre-measured transfer function from the DAC output to the DARM to evaluate the noise contribution. All DAC noise contributions from the Type-A, -B, and -Bp suspensions were evaluated.

Therefore, the noise contribution from the Type-A suspensions was the largest by a factor of 10. The DAC noise from Type-A suspensions is presented in the top panel of Fig.~\ref{fig:NB}. Although it was still lower than the observed sensitivity, DAC noise reduction is necessary in the next observing run. Whitening and de-whitening filters were installed to avoid the DAC noise of the suspensions; the number of the filters will be increased in the future. If necessary, low-noise drivers with 100 times lower driving power will be used when the observed sensitivity approaches the design sensitivity of KAGRA.

\subsubsection{Mains AC power} 
The mains AC power at the KAGRA site (located in the western part of Japan) is 100~V$_{\rm rms}$, 60~Hz, and all electronics, even AC-DC converters, originally operate with the same specifications. 
Therefore, the mains power and its harmonics can appear in any analog circuits as electrical noise, e.g., in PDs and actuators. 
Such noise is reduced to as low as possible by eliminating the AC devices and placing the AC-DC converters in a room away from the interferometer. 

An ADC channel with no connected sensor was used as a witness channel to evaluate its contribution to the DARM signal.
Coherence between the ADC witness channel and DARM signals had clear peaks at $f=60n$~Hz ($n=1,2,3, ...$), of which only $n=1$ was dominant in the DARM (the bottom panel in Fig.~\ref{fig:NB}). We plan to install voltmeters for AC and DC power lines to obtain more detailed noise investigation and mitigation in future observations.

\subsection{Noise contribution between 100~Hz and 400~Hz}

Acoustic noise contributes significantly between 100~Hz and 400~Hz. 
We observed the fundamental violin modes (transverse standing waves of the sapphire fibers that suspended TMs) as a number of peaks.

\subsubsection{Acoustic noise} 

Acoustic vibration originates from many apparatuses in the experimental site including air-conditioners, vacuum pumps, and electrical devices. This acoustic field induces the mechanical vibration of a vacuum chamber and duct, optical table, or in-air optics, and it can contaminate the interferometer signal as scattered light noise or other noise. 
The acoustic field ({\it i.e.}, the sound pressure) $G_\mathrm{mic}(f')$ was measured using microphones located in the experimental site; they are one of the physical environmental monitoring (PEM) systems used to monitor environmental conditions. Details of the KAGRA PEM are reported in Ref.~\cite{Akutsu:2020tpd}. 

The power spectral density of the acoustic noise in the DARM $G_\mathrm{acoustic}(f)$ is evaluated using 

\begin{eqnarray}
 G_\mathrm{acoustic}(f) = \int R(f,f') \times G_\mathrm{mic}(f') df', 
\label{response function}
\end{eqnarray}
where $R(f,f')$ denotes a response function, and $G_\mathrm{mic}(f')$ represents the power spectral density of the microphone signal. This model considers the frequency conversion effect and when the frequency conversion does not occur, the equation is simplified as
\begin{eqnarray}
 G_\mathrm{acoustic}(f) =  C^2(f) \times G_\mathrm{mic}(f), 
\label{coupling function}
\end{eqnarray}
where $C(f)$ denotes a coupling function.

The measurements for the $R(f,f')$ for the PSL room and power-recycling (PR) booth were performed using single line acoustic injection in the post-commissioning term (June 2020). 
In the PR booth, the frequency conversion was confirmed, and the response function model in Eq.~(\ref{response function}) was applied to noise projection. However, in the PSL room, frequency conversion was not observed, and the coupling function model in Eq.~(\ref{coupling function}) was applied.
Further details are described in Ref.~\cite{Washimi:2020slk}. 

The hammering tests for the vacuum systems and optical tables (around the IMC and PRCL) were performed to identify where the acoustic field affected the interferometer signal. 
Finally, the largest contribution to the O3GK sensitivity was identified as the scattered light in the vacuum bellows between the IMC and IFI chambers (Fig.~\ref{fig:ifo}).

The acoustic noise limited the O3GK sensitivity in some frequencies between 100~Hz and 400~Hz.
Therefore, the following approaches are planned to reduce them in the next observation: 
(1) Introduction of soundproofing material at the experimental site to reduce the acoustic field, 
(2) Mitigation of scattered light propagation using new optical baffles and beam dumps inside and outside the vacuum enclosure in addition to the currently installed ones \cite{Akutsu:2016khy},
(3) Acoustic contamination in the DARM signal subtracted by independent component analysis performed in the iKAGRA study~\cite{KAGRA:2019cqm}.

\subsubsection{Violin modes of sapphire fibers}\label{sec:violin}
Peaks observed at approximately 180~Hz, 420~Hz, and 740~Hz include the violin modes, which are the transverse standing waves of sapphire fibers that suspend the TM. The masses of fibers cause back actions on the center of the mirror. A single violin-mode peak can be indicated as 32 peaks in the sensitivity [4 (number of mirrors) $\times$ 4 (number of fibers for each mirror) $\times$ 2 (two degrees of freedom of the transverse vibration for each fiber)]. The number of modes observed can be less than 32, because some of these modes cannot be observed when the mirror motion induced by fiber vibrations is parallel to the mirror surface.
 
Frequencies of violin modes were derived from the equation of elastic fiber motions under tension by gravity \cite{Gonzalez:1994}. The dissipation in the fiber is expected to be dominated by thermoelastic damping because the KAGRA interferometer was at room temperature in O3GK, as described in Sec.~\ref{ss:thermal noise}. Q-values of the violin modes were calculated from the equation of the violin mode motions and the thermoelastic damping formula. The calculated Q-values of the violin modes cannot be lower than the measured value. Q-values are lower When there are other types of dissipation in or around the fiber. 
The parameters of sapphire suspensions used for violin modes calculation are as follows \cite{Akutsu:2020his}: the mass of the TM is 22.8~kg; the length and diameter of a sapphire fiber is 350~mm and 1.6~mm, respectively. The material properties of sapphire are summarized in Table~\ref{table:sapphire}.

\begin{figure}[htbp]\centering
    \includegraphics[width=15cm]{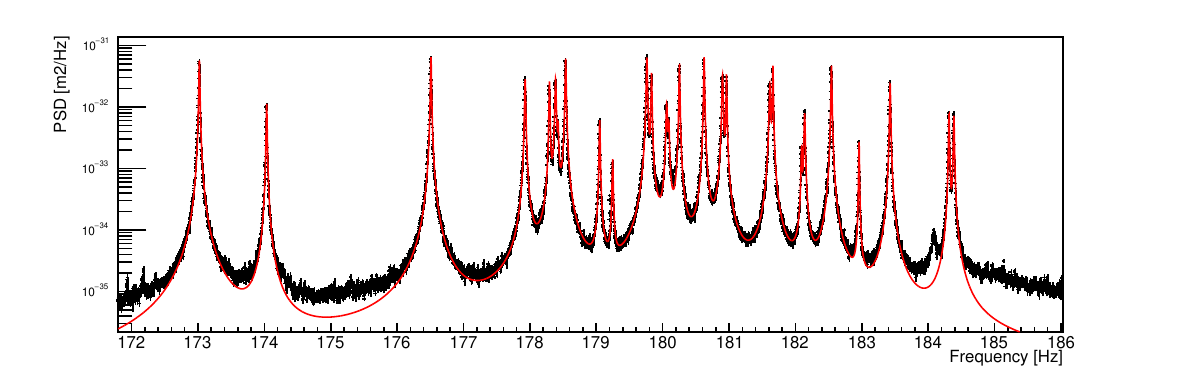}
    \includegraphics[width=15cm]{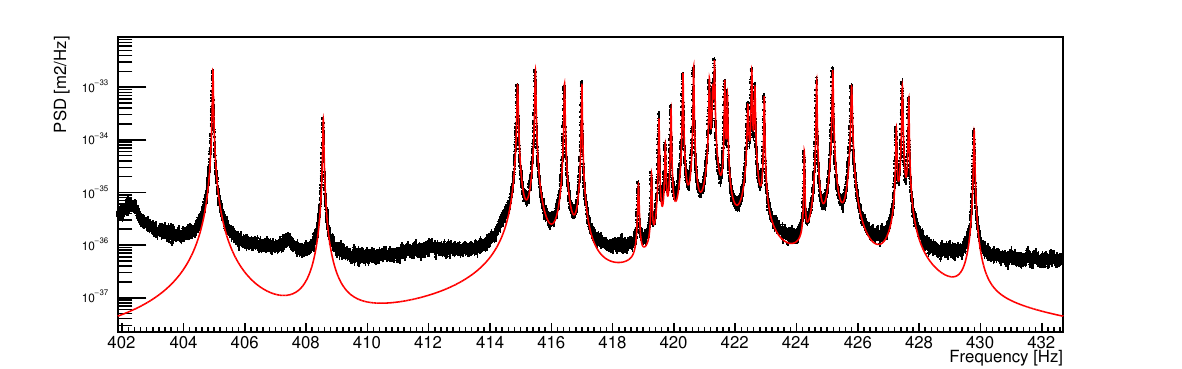}
    \includegraphics[width=15cm]{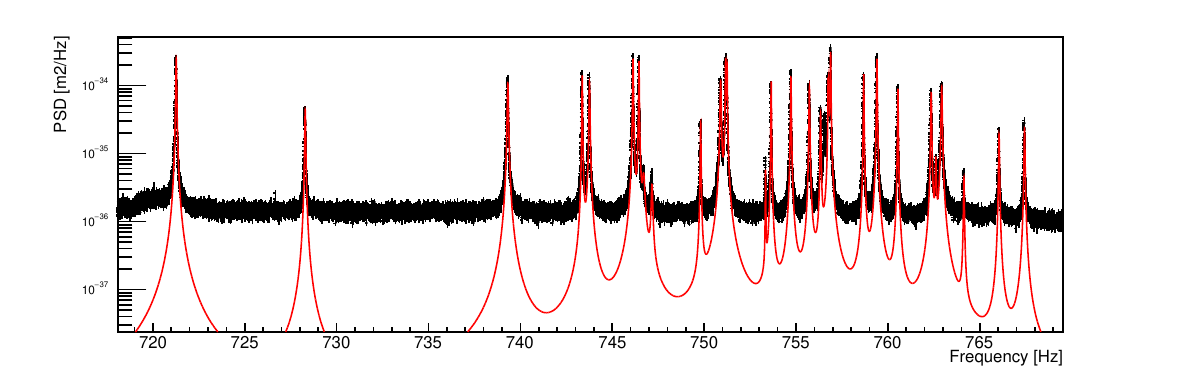}
	\caption{\label{fig:Violin Fitting}Fitting results for the violin mode (top : 1st, middle : 2nd, bottom : 3rd modes) peaks in the power spectrum density (PSD) of the DARM sensitivity. The black and red curves are the observed PSD and fitting result, respectively.}
\end{figure}

From the observation data, the violin mode peaks in Fig.~\ref{fig:Violin Fitting} are evaluated by fitting the power spectral density with
\begin{equation}
    G_{\rm violin}(f) = \sum_i \frac{G_i f_i^2 \varGamma_i^2}{(f^2-f_i^2)^2 +f^2\varGamma_i^2}, 
\end{equation}
where $i$, $G_i$, $f_i$, $\varGamma_i=f_i/Q_i$, and $Q_i$ denote the index for each peak, peak value, peak frequency, FWHM, and Q-value for each peak, respectively.
The results in Table~\ref{table:Violin} are consistent with the theoretical expectation of the violin modes. The number of peaks are less than or equal to 32.
The values of the peak frequencies are listed in the Appendix~\ref{ptep02_linelist}.

Although we expect the excited violin modes observed in O3GK to be thermal fluctuations of the suspension, further investigation is required. If the violin-mode vibration is dominated by thermal fluctuation, the amplitude of the violin modes can be suppressed by cooling. 

\begin{table}[htbp]\centering
\begin{tabular}{l|ccc}
\hline
                         & 1st mode        & 2nd mode        & 3rd mode \\ \hline
Number of identified peaks& 27             &  30             & 31 \\ 
Peak frequency (theory)  & 175~Hz          & 410~Hz          & 732~Hz \\
Peak frequency (mean)    & 180.10~Hz       & 420.84~Hz       & 752.47~Hz \\
Peak frequency (lowest)  & 173.02~Hz       & 404.96~Hz       & 721.27~Hz\\ 
Peak frequency (highest) & 184.39~Hz       & 429.80~Hz       & 767.44~Hz \\
Q-value (theory)         &$2.7 \times 10^4$&$4.7 \times 10^4$&$7.0 \times 10^4$ \\
Q-value (mean)           &$1.30\times 10^4$&$1.80\times 10^4$&$2.12\times 10^4$ \\ 
Q-value (std. dev.)      &$2.0 \times 10^3$& $3.7\times 10^3$&$6.9 \times 10^3$\\ 
\hline
\end{tabular}
\caption{\label{table:Violin}Evaluated parameters of the violin mode peaks.}
\end{table}

\begin{table}[htbp]\centering
\begin{tabular}{l|c}
\hline
Young's modulus &  $4.0\times 10^{11}$~Pa \\
Poisson ratio & 0.29 \\ 
Density  & $4.0\times10^{3}$~kg/m$^3$ \\
Thermal expansion & $5.4\times10^{-6}$~/K \\
Specific heat & $7.73\times10^2$~ J/K/kg\\
Thermal conductivity & 46~W/m/K\\
\hline
\end{tabular}
\caption{\label{table:sapphire}Material properties of sapphire at room temperature.}
\end{table}

\subsubsection{Thermal noise of the sapphire fibers and sapphire mirrors}\label{ss:thermal noise}
The suspended mirrors receive a random energy flow from the heat bath. This random energy flow shifts the fibers suspending the mirrors and deforms the surfaces of the mirrors. The arm length fluctuates because of these motions. They are the fundamental noise sources, and they are called the thermal noise of the suspension and mirror. The details are explained in Ref.~\cite{Akutsu:2020his} and the formulae are briefly explained as follows. 

The fluctuation-dissipation theorem predicts the relationship between the thermal fluctuation and dissipation in the system. A solid exhibits two types of internal dissipation: (i) Thermoelastic damping, which is the thermal relaxation caused by inhomogeneous strain and thermal expansion; the strength of this damping has a frequency dependence. (ii) Structure damping; the damping strength is almost independent of the frequency. The dissipation in the KAGRA TMs at room temperature is dominated by the thermoelastic damping in the substrate and the structure damping in the reflective coating. The thermal noise (power spectral density) caused by thermoelastic damping in a mirror substrate at room temperature is described in Ref.~\cite{Braginsky:1999rp}
\begin{equation}
    G_{\rm substrate} =  \frac{64}{\sqrt{\pi}} {\alpha_{\rm sub}}^2 (1+\sigma_{\rm sub})^2 \frac{k_{\rm B} T^2}{\left(\rho_{\rm sub}C_{\rm sub}\right)^2}\frac{\kappa_{\rm sub}}{{w_0}^3} \frac1{(2 \pi f)^2},
\label{eq:thermoelastic mirror}
\end{equation}
the parameters $\alpha_{\rm sub}$, $\kappa_{\rm sub}$, $C_{\rm sub}$, $\rho_{\rm sub}$, and $\sigma_{\rm sub}$ denote the thermal expansion coefficient, thermal conductivity, specific heat capacity, density, and Poisson ratio of the mirror substrate, respectively. These values are summarized in Table \ref{table:sapphire}. The parameters $f$, $w_{0}$, $k_{\rm B}$, and $T$ denote the frequency, beam radius at the mirror (35~mm \cite{Akutsu:2020his}), Boltzmann constant, and temperature, respectively. For KAGRA's sapphire substrate mirrors in O3GK (room temperature), Eq.~(\ref{eq:thermoelastic mirror}) is rewritten as the amplitude spectral density 
\begin{equation}
\sqrt{G_{\rm substrate}} = 2.5 \times 10^{-20} [{\rm m}/\sqrt{\rm Hz}] \left(\frac{100~{\rm Hz}}{f}\right).
\label{eq:thermoelastic mirror 2}
\end{equation}
The power spectral density of the coating thermal noise is written as (assuming Young's modulus and Poisson ratio of the coating are similar to those of the substrate)
\begin{equation}
G_{\rm coating} =  \frac{16 k_{\rm B}T (1+\sigma_{\rm sub})(1-2\sigma_{\rm sub})(d_{\rm ITM}+d_{\rm ETM})\phi_{\rm coating}}{\pi E_{\rm sub} {w_0}^2 (2 \pi f)},
\label{eq:coating thermal noise}
\end{equation}
where $E_{\rm sub}$, $d_{\rm ITM}$, $d_{\rm ETM}$, and $\phi_{\rm coating}$ denote Young's modulus of the substrate, thickness of coating of ITM (3.7~$\mu$m) and ETM (6.5~$\mu$m), and loss angle of the coatings ($4\times 10^{-4}$), respectively \cite{Akutsu:2020his}. For KAGRA O3GK, Eq.~(\ref{eq:coating thermal noise}) can be rewritten in the amplitude spectral density as 
\begin{equation}
\sqrt{G_{\rm coating}} = 2.2 \times 10^{-20} [{\rm m}/\sqrt{\rm Hz}] \left(\frac{100~{\rm Hz}}{f}\right)^{1/2}.
\label{eq:coating thermal noise 2}
\end{equation}
The simplified formula, Eq.~(\ref{eq:coating thermal noise}), underestimates (1.8 times) the coating thermal noise because of the differences in the elastic properties between the sapphire and coating. Equation~(\ref{eq:coating thermal noise 2}) considers this difference between the substrate and coating~\cite{Crooks:2002}.
For simplicity, Fig.~\ref{fig:NB} shows the square root of quadrature summation of Eqs.~(\ref{eq:thermoelastic mirror 2}) and (\ref{eq:coating thermal noise 2}) as the yellow curve; this summation did not limit O3GK sensitivity.

The dissipation in the TM's suspension is dominated by thermoelastic damping in their sapphire fibers. In this case, the dissipation can be derived from the theory of elasticity and the elastic and thermal properties of the sapphire fibers. 
A simplified expression of the power spectral density of the thermal noise in its pendulum mode is
\begin{eqnarray}
G_{\rm pendulum} &=&  \frac{16k_{\rm B}Tg}{m (2 \pi f)^5\ell_{\rm fiber}}\phi_{\rm fiberthermo}\times\sqrt{\frac{4\pi E_{\rm fiber}}{mg\ell_{\rm fiber}^2}}\left(\frac{d_{\rm fiber}}{4}\right)^2,\label{eq:suspTN}\\
\phi_{\rm fiber thermo} &=& \frac{\alpha_{\rm fiber}^2 E_{\rm fiber} T}{\rho_{\rm fiber} C_{\rm fiber}}\frac{f/f_0}{1+(f/f_0)^2},\\
f_0 &=& 2.16 \frac{\kappa_{\rm fiber}}{\rho_{\rm fiber} C_{\rm fiber} d_{\rm fiber}^2}, 
\end{eqnarray}
where $g$ and $m$ denote the gravitational acceleration and the mass of the TM, respectively; $\ell_{\rm fiber}$, $E_{\rm fiber}$, and $d_{\rm fiber}$ denote the length, Young's modulus, and diameter of the sapphire fiber, respectively; $\phi_{\rm fiberthermo}$ denotes the loss angles by thermoelastic damping of the fiber; and $\alpha_{\rm fiber}$, $\rho_{\rm fiber}$, $C_{\rm fiber}$, and $\kappa_{\rm fiber}$ denote the thermal expansion coefficient, density, specific heat capacity, and thermal conductivity of the fibers, respectively.

This simplified formula is appropriate when the fibers are thin. In the case of KAGRA, it is not true because the fiber must be thick to extract the heat generated in the mirror. The elasticity of the sapphire fibers are considered as shown in Ref.~\cite{Gonzalez:1994}. 
For the KAGRA sapphire suspension at room temperature, the amplitude power spectrum between 10~Hz and 100~Hz is approximated as follows ($f_0$ is approximately 13~Hz) 
\begin{equation}
\sqrt{G_{\rm pendulum}} = 6.0 \times 10^{-18} [{\rm m}/\sqrt{\rm Hz}] \left(\frac{30~{\rm Hz}}{f}\right)^3 .
\end{equation}
The violin modes appear above 100~Hz.
The suspension thermal noise derived by these formulae is presented in the gray line in the top panel of Fig.~\ref{fig:NB}, which indicates that the suspension thermal noise did not limit O3GK sensitivity. Around 100~Hz, the thermal noise was approximately ten times lower than the O3GK sensitivity. For future operations, thermal noise reduction by cooling is inevitable. The thermal noise at 100~K is approximately one order of magnitude lower than at room temperature \cite{Akutsu:2020his}. The thermal noise is approximately three times lower than that at 100~K when the temperature is 22~K (the KAGRA specification value).  

\medskip

\subsection{Noise contribution above 400~Hz}

Above 400~Hz, the main noise contributions originated from the light. Between 400~Hz and 2~kHz, the shot noise dominated the sensitivity. Above 2~kHz, the laser frequency noise contributed significantly.

\subsubsection{Shot Noise}
\label{sec:shot_noise}

The quantum fluctuation of the photon number at a PD (DC PD in Fig.~\ref{fig:ifo}) is called shot noise. We evaluated the shot noise using two different formulae (an experimental and a theoretical one), and we confirmed that the results are consistent. The result from the theoretical formula is shown in Fig.~\ref{fig:NB}. The experimental formula includes the measured interferometer response. For the theoretical formula, the interferometer response is represented by the optical gain and cavity pole frequency. The experimental formula of the power spectral density of the mirror displacement equivalent noise caused by shot noise is expressed as 
\begin{equation}
G_{\rm shot} = \frac{2 h c P_{\rm DC} }{\lambda |H_{\rm int}|^2},\label{eq:shot noise2}
\end{equation}
where $H_{\rm int}$ denotes the interferometer response from the mirror displacement to the light power change at the PD. $h, c, P_{\rm DC}$, and $\lambda$, represent the Planck constant, speed of light, average DC component of the power at the PD, and wavelength of the light, respectively. We derived the shot noise from the measured values, $P_{\rm DC}$ and $H_{\rm int}$. 
Values $P_{\rm DC}$ and $H_{\rm int}$ depend on the operation condition of the interferometer (amount of the local oscillator field and the contrast defect). The theoretical shot noise power spectral density $G_{\rm shot}$ is represented when the amplitude of the local oscillator is sufficiently larger than the contrast defect as follows 
\begin{eqnarray}
G_{\rm shot}&=&\frac{h \lambda c}{32 {\cal G_{\rm p}} P \delta {\cal F}^2}\left[1+\left(\frac{f}{f_{\rm cut}}\right)^2\right],\label{eq:shot noise}\\   
f_{\rm cut}&=& \frac{c}{4L {\cal F}}.\label{eq:cut off}
\end{eqnarray}
Parameters ${\cal G_{\rm p}}, P, \delta, L$, and ${\cal F}$ denote PR gain, injected light power at the PRM, optical power transmittance between the BS and DC PD in Fig.~\ref{fig:ifo}, and the length and finesse of the arm cavities, respectively.  
The averaged value of the measured finesse of the two 3-km arm cavities is $1.4\times 10^3$.
The measured finesse implies that the cut-off frequency is 18 Hz, and the amplitude spectral density of the shot noise in O3GK is represented by
\begin{equation}
\label{eq:sn2}
\sqrt{G_{\rm shot}}= 3.8 \times 10^{-20} [{\rm m}/ \sqrt{\rm Hz}] \left[1+\left(\frac{f}{\rm 18 ~Hz}\right)^2\right]^{1/2},
\end{equation}
where ${\cal G_{\rm p}}=11$, $P=5.3~$W, and $\delta=0.04$. The estimated shot noise level derived from Eq.~(\ref{eq:sn2}) is depicted as a black dashed curve in the upper plot of Fig.~\ref{fig:NB}. The shot noise limited the observed sensitivity between 400~Hz and 2~kHz. 
Only 4\% of the interfered light at the BS ($\delta$=0.04) arrived at the PD. We investigated what causes such a low optical transmittance. The SRM (details are in caption of Fig.~\ref{fig:ifo}) transmittance is 30\%. The total transmittance of OFI and OMC was 36\%. In addition, unknown transmittance reduction (approximately 40\%) was observed. The total transmittance (products of transmittance of each part) is 4\%.   

We will proceed to increase the optical transmittance between the BS and DC PD. The shot noise amplitude is expected to be reduced by a factor of 4-5 when the transmittance is close to unity. Further, we plan to introduce a higher power laser (more than 50 W at the BS) and resonant sideband extraction (RSE, details are provided in Ref.~\cite{Akutsu:2020his,Mizuno:1993cj}) in the near future. For the latter, it is necessary to align the SRM.

\subsubsection{Radiation Pressure Noise}
\label{sec:radiation pressure noise}

Radiation pressure noise is caused by the quantum fluctuation of the photon number at the test masses. The sapphire main mirrors receive a back action force by photons in the laser light; the photon number fluctuation changes the force and position of the sapphire main mirrors. The power spectral density of the radiation pressure noise in the mirror displacement is expressed as 
\begin{equation}
G_{\rm radiation} = \left(\frac{2}{m (2 \pi f)^2} \frac{2{\cal F}}{\pi}\right)^2\frac{8 h {\cal G_{\rm p}}P}{c \lambda}\left[1+\left(\frac{f}{f_{\rm cut}}\right)^2\right]^{-1},    
\end{equation}
where $m$ denotes the mass of the TM. The calculated amplitude spectral density for the KAGRA detector in O3GK is 
\begin{equation}
\sqrt{G_{\rm radiation}} = 6.2 \times 10^{-17}  [{\rm m}/ \sqrt{\rm Hz}] \left(\frac{1~{\rm Hz}}{f}\right)^2 \left[1+\left(\frac{f}{18~{\rm Hz}}\right)^2\right]^{-1/2}.   
\end{equation}
This noise did not limit O3GK sensitivity as are shown in Fig.~\ref{fig:NB}.
While the radiation pressure noise increases with laser power, it decreases with the introduction of the SRM in the RSE configuration. Consequently, the radiation pressure noise in design sensitivity and that in O3GK observation are comparable.


\subsubsection{Laser frequency noise}
\label{sec:laser_freq}
A Michelson interferometer is a precise length measurement device with the wavelength as a ruler. The fluctuation of the wavelength (or frequency) is the measurement limit of the interferometer. In an ideal case, the laser frequency noise does not contaminate the GW signal because its effect is common to the two arms and is canceled in the DARM. In contrast, the frequency fluctuation appears in the CARM. The CARM signal is used as the reference for frequency stabilization because CARM does not include any GW signals and the 3-km arm cavities are the longest and most stable frequency references in the KAGRA interferometer.

In the actual senario, the above assumption is not perfectly accurate and there is coupling from CARM to DARM because of the asymmetry of both arms, such as by the differences in the length and finesse. 
The amplitude spectral density of the CARM error signal was detected and the transfer function from the CARM error signal to the DARM was measured to evaluate the frequency noise contribution. The laser frequency noise contribution in the DARM is represented by the product of the amplitude spectral density and transfer function. The calculated laser frequency noise was dominant above 2~kHz in O3GK as presented as a light green curve in the upper plot of Fig.~\ref{fig:NB}. 

The coupling between CARM and DARM was measured and observed to be larger than the  expected level derived using the simulation model that included known or assumed differences in the transmittance of the ITMs, finesse of the arm cavities (as depicted in Table \ref{table:param}), reflected light phase map, and transmission wavefront error map that essentially displays the non-uniformity of the optical thickness of the mirror \cite{Somiya:2019TWE}. A possible cause of the larger CARM-DARM coupling is the birefringence of the test mass mirrors. Further studies on birefringence are in progress.
A lower noise in the CARM error signal and a higher servo gain for frequency stabilization are necessary for a better frequency noise level. Below 1 kHz, the laser frequency noise is limited by the technical noise of the control loop (such as dark current noise) of the PD to obtain the CARM error signal. 
The technical noise contribution decreases if the light power at the PD increases because the CARM error signal is proportional to the light power while the technical noise is independent of the light power. The signal-to-noise ratio can improved for several times compared to the current level by increasing the laser power at the PD for the frequency stabilization in the future.
Further, installing a vacuum system for the CARM PD will help reduce the noise caused by air turbulence on the laser beam.
The gain of the servo filter needs further optimization in the high-frequency region above 1~kHz to suppress the frequency noise further.

\medskip

\subsubsection{Laser intensity noise}
\label{sec:laser_intensity}
There is an intentional offset in the light power at the PD (DC PD in Fig.~\ref{fig:ifo}) for the DC readout to extract GW signals. Further, the Michelson interferometer asymmetry introduces some laser power to the PD. Here, the intensity change imprinted by the GW signal cannot be distinguished from the laser intensity fluctuations. In KAGRA, the intensity is monitored using an in-air PD that receives the transmitted beam through IMMT1 shown in Fig.~\ref{fig:ifo} to suppress the laser intensity noise. The monitored intensity is used to adjust the injected power to the interferometer by using an acousto-optic modulator (AOM) installed in the PSL system. Nevertheless, some residual fluctuations still exist. 

The transfer function from the error signal of the intensity monitor refers to the DARM to evaluate the intensity noise contribution, and the intensity fluctuation in this error signal is measured. The product of the transfer function and the fluctuation is the intensity noise contribution to the DARM.

Although the calculated intensity noise contribution did not limit the sensitivity of O3GK, further reduction is necessary. The reduction strategies of the intensity stabilization system include reducing the coupling from the intensity fluctuation to DARM, and suppressing the intensity fluctuation. For the former, the interferometer asymmetry must be as low as possible, like in the case of the laser frequency noise \cite{Michimura:2021LVK}.
For the latter, one serious issue includes the intensity fluctuation at IMMT1 without intensity stabilization to be higher than that of the laser source. Therefore, there must be noise sources between the laser source and the IMMT1 transmission, and these must be removed.
Further, the introduction of a higher-power laser will contribute to a lower intensity fluctuation. A higher power is useful in suppressing the shot noise in the intensity fluctuation monitor. The number of PDs will be increased from two to four to receive higher monitoring power.
Another noise source in the intensity monitor could be the beam jitter. We found that the injected beam jitter caused fluctuations in the error signal obtained at the intensity monitoring PD. The possible reasons are the inhomogeneous quantum efficiency on the receiving area and beam position vibration of the photodiodes. We plan to install a jitter stabilization system with quadrant PDs and piezo actuators placed before the intensity monitoring PDs to mitigate the beam jitter. We plan to place the optics for the intensity stabilization system in a vacuum chamber to further suppress intensity monitor noise. 

\subsubsection{OMC PD dark noise}

The OMC is a bow-tie shaped cavity in Fig.~\ref{fig:ifo} used to filter unwanted RF sideband fields and higher-order spatial modes generated by the interferometer imperfections.
The DARM signals containing the GW signals are extracted as the change in the transmitted power through the OMC. 
The signal of the PD receiving this light contains noises from the ambient light fluctuations, photodiode dark current, electric circuit noises of the transimpedance amplifier, noise of the ADC, and so on. The sum of those background noises is called the ``OMC PD dark noise.''

The PD output was measured without laser light to evaluate this noise. This was performed after the observation run and was converted to the DARM-equivalent value with calibration parameters used for the observation run, assuming that dark-current noise and electrical noise did not drastically change between the observation period and the measurement.
This dark noise was discovered to be lower than the observed noise in O3GK as shown in the light gray curve in the upper plot in Fig.~\ref{fig:NB}. 

In future observing runs,
the signal-to-noise ratio of the PD will be improved
by stronger signals with a higher laser power of the input beam.

\subsection{Injected lines}
\label{sec:lines}

Sinusoidal signals (lines) were intentionally injected to monitor the status of the interferometer. 
These injected lines are shown in the bottom panel of Fig.~\ref{fig:NB}; they are discussed here even though some of them were not used during the observation but appeared in the sensitivity.
The list of line frequencies is provided in the Appendix~\ref{ptep02_linelist}.

\subsubsection{Alignment dither system (ADS)}
IMMT2, BS, and PR3 were excited with angular degrees of freedom for the ADS. In addition, SR3 was excited and controlled for the green beam alignment to maximize the arm transmission of the green beam. For the same purpose, PR3 excitation was used for the green beam alignment.
As represented by the dark green lines in the bottom panel of Fig. 4, the angular excitation described in Sec.~\ref{sec:asc} was coupled to the sensitivity because of the imperfect alignment or mis-centered beam spots. The excitation frequencies were 20--45 Hz.

\subsubsection{Beam Position Centering (BPC)}
BPC was intended; however, no BPC control loops were used during the observing run due to technical reasons.
The pitch and yaw excitation applied to the input and end mirrors persisted and coupled into the sensitivity, as represented by the light green lines in the bottom plot of Fig.~\ref{fig:NB} at approximately 1 kHz, which were selected to avoid the mechanical resonance of the suspensions.

\subsubsection{Calibration lines for DARM displacement} 
During the observation, the calibration lines are injected in two ways to monitor the time-dependence of the calibration factor of DARM displacement. 
One is through the coil magnet actuators for controlling the positions and angles of each end test mass mirrors, with the frequencies of 28.3~Hz and 30.7~Hz for ETMX and ETMY, respectively. 
The other is through the photon calibrator~\cite{Akutsu:2020tpd} for the ETMX mirror, with frequencies of 29.5~Hz, 79.7~Hz, and 859.7~Hz.

\subsubsection{Calibration lines for MICH/PRCL Feed-forward} 
MICH and PRCL were excited at 66.6~Hz and 63.1~Hz, respectively, to monitor the time-dependence of the transfer function from them to DARM. 
These lines can be used to calibrate the transfer functions in the offline analysis.

\subsubsection{Reducing the effects of injection lines}
The ADS or BPC lines will not be used once global alignment control \cite{Fritschel:1998ctz} is implemented in future observation runs.
However, the calibration lines cannot be removed at the hardware level because of their functions; however, they can be subtracted at software processing such as during DARM reconstruction or some offline analysis because the actuation powers are also recorded.

\section{Future Prospects} \label{ptep02_sec4}

The noise contributions in the KAGRA detector during the O3GK run were studied, and the noise mitigation strategies were discussed. The new challenges and issues to be expected in the near future are summarized. 

The optical configuration will be significantly upgraded from the power-recycled Fabry--P\'{e}rot--Michelson interferometer to the RSE which has an additional SRC at the AS port in next observing runs. This will help broaden the sensitive bandwidth for GW signals. A high-power laser source is planned to be installed for reducing the shot noise level because the high-frequency region of RSE sensitivity will be still limited by the shot noise, similar to the current case.

Another major upgrade will be lowering the thermal noise drastically by performing a cryogenic operation. Investigations to cool down without problems (for example, ice layer on mirror surface~\cite{Akutsu:2020his}) are in progress. 

Commissioning of the high-power operation is challenging.
Interferometer mirrors absorb the laser power,
and hot spots are created in the optics when the input power is high.
Further, hot spots distort the wavefronts of the laser beam through
local refractive index changes and the thermal expansion of the mirror material~\cite{RyanChristopher:2003zdq}.
However, these thermal effects are negligible for the cryogenic sapphire mirrors \cite{Tomaru:2002qz}. Therefore, sapphire mirrors must be cooled prior to the power increase~\cite{Degallaix:2012}. 
In addition, because the thermal aberration effects in room temperature optics (e.g., beamsplitter) are experimentally unknown for the KAGRA detector, we must investigate these effects in order to identify its noise behavior at room temperature.
If such thermal aberration effects are problematic for the interferometer operation, active thermal compensation systems \cite{aLIGO:2016yis} maybe necessary for the future.
Other issue related to the high-power laser is the parametric instability \cite{Braginsky:2001} that causes the unstable sapphire mirror elastic vibration and excitation of higher optical modes in the 3-km cavities. Although the number of unstable modes is an order magnitude less than those of LIGO and Virgo \cite{Yamamoto:2008}, the suppression system for the cooled sapphire mirrors is necessary and being studied.

There remain a few unsolved issues for future commissioning. An unexpected large coupling from laser frequency and intensity noise to the sensitivity was observed in O3GK, and it is likely to remain because of large asymmetries between the X and Y arm cavities. Asymmetries between the two arms may be responsible for the couplings from the auxiliary degrees of freedom (as presented in Sec.~\ref{sec:MICH/PRCL noise}). As discussed in Sec.~\ref{sec:laser_freq}, the birefringence of the sapphire substrate is a possible cause of the asymmetries.
Further investigation to characterize the birefringence effect is in progress.

In the upgraded detector, noise couplings may change and have to be re-evaluated. Noise sources that are not observed in O3GK may be complicated at a better sensitivity. According to the experiences in Advanced LIGO and Advanced VIRGO, scattering light noise and angle-to-length coupling noise may appear in a broad frequency range and between 10~Hz to 100~Hz, respectively. Additional optical baffles and beam dumps are being prepared for various places in the vacuum enclosure and on the optical tables in the air to mitigate the scattered light.
The angle-to-length coupling noise problem will be addressed by a new global angular sensing and control scheme for the KAGRA detector.

The KAGRA noise was reduced by 4.5 orders of magnitude in O3GK compared with the first cryogenic operation in 2018~\cite{Akutsu:2019rba}. The measured sensitivity can be explained by adding up the effect of each noise, and the noise reduction strategies are discussed. Based on these studies, the KAGRA sensitivity will be improved for the next observing run aiming to contribute to the GW astrophysics.


\section*{Acknowledgment}
This work was supported by MEXT, JSPS Leading-edge Research Infrastructure Program, JSPS Grant-in-Aid for Specially Promoted Research 26000005, JSPS Grant-in-Aid for Scientific Research on Innovative Areas 2905: JP17H06358, JP17H06361 and JP17H06364, JSPS Core-to-Core Program A. Advanced Research Networks, JSPS Grant-in-Aid for Scientific Research (S) 17H06133 and 20H05639 , JSPS Grant-in-Aid for Transformative Research Areas (A) 20A203: JP20H05854, the joint research program of the Institute for Cosmic Ray Research, University of Tokyo, National Research Foundation (NRF), the Mitsubishi Foundation, Computing Infrastructure Project of KISTI-GSDC in Korea, Academia Sinica (AS), AS Grid Center (ASGC) and the Ministry of Science and Technology (MoST) in Taiwan under grants including AS-CDA-105-M06, Advanced Technology Center (ATC) of NAOJ, Mechanical Engineering Center of KEK, the LIGO project, the Virgo project, and Terri Pearce. Finally, we would like to thank all of the essential workers in the KAGRA observatory; we would not have been able to complete this work without them.

\bibliography{reference}

\newpage
\appendix
\section{List of the identified lines} \label{ptep02_linelist}

\begin{table}[htbp]\centering
\begin{tabular}{ll}
\hline
Frequency [Hz] & Type \\ \hline 
15.30&	ADS (SR3, Yaw) \\
20.99&	ETMX eigenmode (Vertical, RM-chain 1st) \\
21.36&	ITMY eigenmode (Vertical, RM-chain 1st) \\
22.10&	ADS (IMMT2, Yaw) \\  
23.39&	ETMX eigenmode (Roll, TM-chain 1st) \\
23.39&	ETMY eigenmode (Roll, TM-chain 1st) \\
23.39&	ITMX eigenmode (Roll, TM-chain 1st) \\
23.54&	ITMY eigenmode (Roll, TM-chain 1st) \\ 
23.83&	ETMY eigenmode (Vertical, RM-chain 1st) \\
25.40&	ADS (SR3, Pitch) \\
26.10&	ADS (BS, Yaw) \\
29.50&	DARM Calibration (ETMX PCal) \\
28.30&	DARM Calibration (ETMX actuator) \\
30.70&	DARM Calibration (ETMY actuator) \\
30.10&	ADS (PR3, Yaw) \\                                                              
38.10&	ADS (IMMT2, Pitch) \\
39.75&	ITMY eigenmode (Transverse,  TM-chain) \\
39.36&	Type-A eigenmode (Vertical, TM-chain) \\
39.75&	ITMY eigenmode (Vertical, TM-chain) \\
40.13&	Type-A eigenmode (Vertical, TM-chain) \\
42.10&	ADS (BS, Pitch) \\
45.23&	ETMX eigenmode (Pitch, TM-chain 2nd) \\
45.48&	ITMY eigenmode (Pitch, TM-chain 2nd) \\
46.07&	ITMX eigenmode (Pitch, TM-chain 2nd) \\
46.10&	ADS (PR3, Pitch) \\
54.19&	ETMX eigenmode (Roll, TM-chain  2nd) \\
54.33&	ETMY eigenmode (Roll, TM-chain  2nd) \\
54.51&	ITTM eigenmode (Roll, TM-chain  2nd) \\ 
55.27&	ITMX eigenmode (Roll, TM-chain  2nd) \\
60.00&	AC power line \\   
63.10&	PRCL feed foword \\ 
66.60&	MICH feed foword \\   
79.70&	DARM Calibration (ETMX PCal) \\ \hline
\end{tabular}
\caption{\label{table:line}List of the identified lines.}
\end{table}

\begin{table}[htbp]\centering
\begin{tabular}{ll}
\hline
Frequency [Hz] & Type \\ \hline 
173.021&	Violin mode (1st) of a cryo-payload \\
174.035&	Violin mode (1st) of a cryo-payload \\
176.510&	Violin mode (1st) of a cryo-payload \\
177.929&	Violin mode (1st) of a cryo-payload \\
178.295&	Violin mode (1st) of a cryo-payload \\ 
178.387&	Violin mode (1st) of a cryo-payload \\
178.425&	Violin mode (1st) of a cryo-payload \\
178.539&	Violin mode (1st) of a cryo-payload \\
179.052&	Violin mode (1st) of a cryo-payload \\ 
179.247&	Violin mode (1st) of a cryo-payload \\
179.762&	Violin mode (1st) of a cryo-payload \\
179.833&	Violin mode (1st) of a cryo-payload \\
180.061&	Violin mode (1st) of a cryo-payload \\
180.093&	Violin mode (1st) of a cryo-payload \\
180.253&	Violin mode (1st) of a cryo-payload \\
180.625&	Violin mode (1st) of a cryo-payload \\
180.902&	Violin mode (1st) of a cryo-payload \\
180.962&	Violin mode (1st) of a cryo-payload \\ 
181.615&	Violin mode (1st) of a cryo-payload \\
181.661&	Violin mode (1st) of a cryo-payload \\
182.093&	Violin mode (1st) of a cryo-payload \\
182.139&	Violin mode (1st) of a cryo-payload \\
182.543&	Violin mode (1st) of a cryo-payload \\
182.959&	Violin mode (1st) of a cryo-payload \\
183.427&	Violin mode (1st) of a cryo-payload \\
184.314&	Violin mode (1st) of a cryo-payload \\
184.387&	Violin mode (1st) of a cryo-payload \\
404.961&	Violin mode (2nd) of a cryo-payload \\
408.562&	Violin mode (2nd) of a cryo-payload \\
414.901&	Violin mode (2nd) of a cryo-payload \\
415.482&	Violin mode (2nd) of a cryo-payload \\
416.433&	Violin mode (2nd) of a cryo-payload \\
416.998&	Violin mode (2nd) of a cryo-payload \\
418.844&	Violin mode (2nd) of a cryo-payload \\
419.270&	Violin mode (2nd) of a cryo-payload \\
419.512&	Violin mode (2nd) of a cryo-payload \\
419.721&	Violin mode (2nd) of a cryo-payload \\
419.906&	Violin mode (2nd) of a cryo-payload \\\hline
\end{tabular}
\caption{\label{table:line}List of the identified lines.}
\end{table}

\begin{table}[htbp]\centering
\begin{tabular}{ll}
\hline
Frequency [Hz] & Type \\ \hline 
420.296&	Violin mode (2nd) of a cryo-payload \\
420.644&	Violin mode (2nd) of a cryo-payload \\
421.157&	Violin mode (2nd) of a cryo-payload \\
421.269&	Violin mode (2nd) of a cryo-payload \\
421.325&	Violin mode (2nd) of a cryo-payload \\
421.659&	Violin mode (2nd) of a cryo-payload \\
421.744&	Violin mode (2nd) of a cryo-payload \\
422.415&	Violin mode (2nd) of a cryo-payload \\
422.540&	Violin mode (2nd) of a cryo-payload \\
422.642&	Violin mode (2nd) of a cryo-payload \\
422.949&	Violin mode (2nd) of a cryo-payload \\
424.255&	Violin mode (2nd) of a cryo-payload \\
424.658&	Violin mode (2nd) of a cryo-payload \\
425.184&	Violin mode (2nd) of a cryo-payload \\
425.798&	Violin mode (2nd) of a cryo-payload \\
427.256&	Violin mode (2nd) of a cryo-payload \\
427.454&	Violin mode (2nd) of a cryo-payload \\
427.662&	Violin mode (2nd) of a cryo-payload \\
429.796&	Violin mode (2nd) of a cryo-payload \\
520.00&	BPC (PRM, Pitch) since April 18 \\
550.00&	BPC (PR2, Yaw) since April 18 \\
580.00&	BPC (PR3, Pitch) since April 18 \\
610.00&	BPC (PRM, Yaw) since April 18 \\
640.00&	BPC (PR2, Pitch) since April 18 \\
670.00&	BPC (PR3, Yaw) since April 18 \\
721.271&	Violin mode (3rd) of a cryo-payload \\
728.291&	Violin mode (3rd) of a cryo-payload \\
739.319&    Violin mode (3rd) of a cryo-payload \\
743.367&	Violin mode (3rd) of a cryo-payload \\
743.762&	Violin mode (3rd) of a cryo-payload \\
746.132&	Violin mode (3rd) of a cryo-payload \\
746.455&	Violin mode (3rd) of a cryo-payload \\
746.707&	Violin mode (3rd) of a cryo-payload \\
747.163&	Violin mode (3rd) of a cryo-payload \\
749.812&	Violin mode (3rd) of a cryo-payload \\\hline
\end{tabular}
\caption{\label{table:line}List of the identified lines.}
\end{table}

\begin{table}[htbp]\centering
\begin{tabular}{ll}
\hline
Frequency [Hz] & Type \\ \hline 
750.900&	Violin mode (3rd) of a cryo-payload \\
751.171&	Violin mode (3rd) of a cryo-payload \\
751.249&	Violin mode (3rd) of a cryo-payload \\
753.344&	Violin mode (3rd) of a cryo-payload \\
753.651&	Violin mode (3rd) of a cryo-payload \\
754.724&	Violin mode (3rd) of a cryo-payload \\
755.729&	Violin mode (3rd) of a cryo-payload \\
756.340&	Violin mode (3rd) of a cryo-payload \\
756.750&	Violin mode (3rd) of a cryo-payload \\
756.786&	Violin mode (3rd) of a cryo-payload \\
756.885&	Violin mode (3rd) of a cryo-payload \\
758.686&	Violin mode (3rd) of a cryo-payload \\
759.405&	Violin mode (3rd) of a cryo-payload \\
760.548&	Violin mode (3rd) of a cryo-payload \\
762.359&	Violin mode (3rd) of a cryo-payload \\
762.611&	Violin mode (3rd) of a cryo-payload \\
762.930&	Violin mode (3rd) of a cryo-payload \\
764.139&	Violin mode (3rd) of a cryo-payload \\
766.052&	Violin mode (3rd) of a cryo-payload \\
767.439&	Violin mode (3rd) of a cryo-payload \\ 
859.70&	DARM Calibration (ETMX PCal) \\
980.00&	BPC (ETMY, Pitch) \\
1010.00&	BPC (ETMX, Yaw) \\ 
1040.00&	BPC (ITMY, Pitch) \\
1070.00&	BPC (ITMX, Yaw) \\
1100.00&	BPC (ETMY, Yaw) \\
1130.00&	BPC (ETMX, Pitch) \\
1160.00&	BPC (ITMY, Yaw) \\
1190.00&	BPC (ITMX, Pitch) \\
\hline
\end{tabular}
\caption{\label{table:line}List of the identified lines.}
\end{table}

\end{document}